\documentclass{article}

\newif\ifpaper
\papertrue

\usepackage{cp_shared/arxiv}
\usepackage[style=authoryear,doi=true,isbn=true,url=false,eprint=true]{biblatex}

\usepackage[utf8]{inputenc} 
\usepackage[T1]{fontenc}    
\usepackage{hyperref}       
\usepackage{url}            
\usepackage{booktabs}       
\usepackage{amsfonts}       
\usepackage{nicefrac}       
\usepackage{microtype}      
\usepackage{graphicx}

\usepackage{doi}

\usepackage{gensymb} 
\usepackage{rotating} 
\usepackage{caption}
\usepackage{subcaption}
\usepackage[nobottomtitles*]{titlesec}
\usepackage{amsmath}
\usepackage{setspace}

\newcommand{\code}[1]{\lstinline[language=bash, basicstyle=\ttfamily\small]|#1|}
\usepackage{color}
\definecolor{lightgrey}{rgb}{0.975, 0.975, 0.975}
\definecolor{midgrey}{rgb}{0.6, 0.6, 0.6}
\definecolor{deepgrey}{rgb}{0.2, 0.2, 0.2}
\definecolor{codegreen}{rgb}{0, 0.7, 0}
\definecolor{codepink}{rgb}{0.8196, 0.10, 0.654}
\usepackage{listings}
\title{Network centrality measures and their correlation to mixed-uses at the pedestrian-scale}

\date{} 

\author{
	\href{https://orcid.org/0000-0003-3790-0638}{
		\includegraphics[width=0.25cm, height=0.25cm, keepaspectratio]{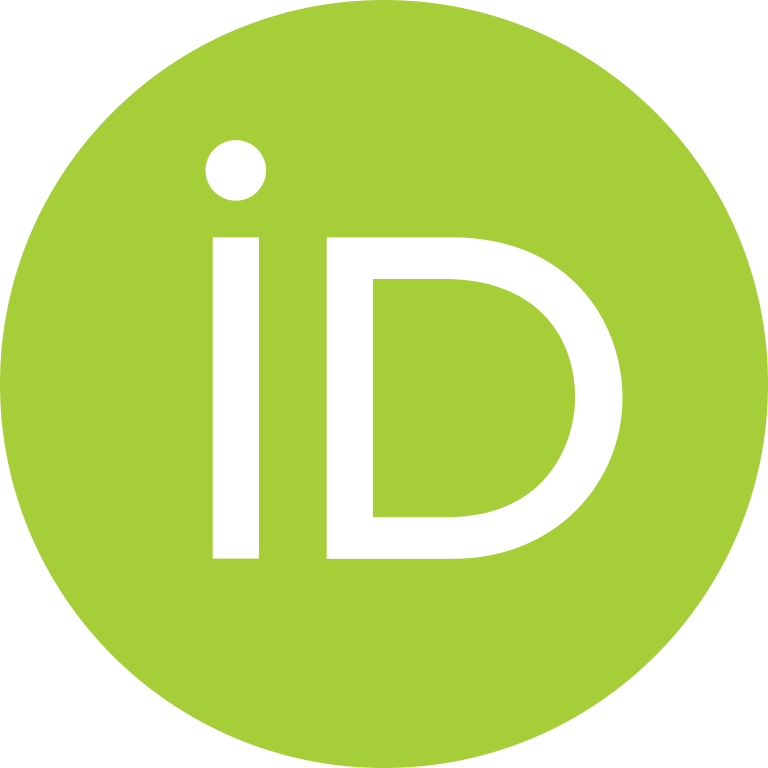}
		\hspace{1mm}Gareth D. Simons
	}
	\thanks{Benchmark Urbanism \texttt{gareth@benchmarkurbanism.com}}
}

\renewcommand{\shorttitle}{Centrality methods at the pedestrian scale}

\hypersetup{
	pdftitle={
		Network centrality measures and their correlation to mixed-uses at the pedestrian-scale
	},
	pdfsubject={
		physics.soc-ph
	},
	pdfauthor={
		Gareth~D.~Simons,
	},
	pdfkeywords={
		computation,
		data science,
		land-use analysis,
		morphometrics,
		network analysis,
		spatial analysis,
		urban analytics,
		urban planning,
		urban morphology,
		urbanism
	},
}

\begin{document}
\maketitle
\begin{abstract}
	Street network analysis holds appeal as a tool for the assessment of pedestrian connectivity and its relation to the intensity and mix of land-uses; however, application within urban-design triggers a range of questions on implementary specifics due to a variety of theories, methods, and considerations and it is not immediately clear which of these might be the most applicable at the pedestrian scale in relation to land-uses. It is, furthermore, difficult to directly evaluate differing approaches on a like-for-like basis without recourse to the underlying algorithms and computational workflows. To this end, the \code{cityseer-api} \code{Python} package is here used to develop, compute, and compare a range of centrality methods which are then applied to the \emph{Ordnance Survey} \emph{Open Roads} dataset for Greater London. The centralities are correlated to high-resolution land-use and mixed-used measures computed from the \emph{Ordnance Survey} \emph{Points of Interest} dataset for the same points of analysis using a spatially precise methodology based on network distances to premise locations.

The comparisons show that mixed-uses correlate more strongly against closeness than betweenness centralities; segmented measures tend to offer slightly stronger correlations than node-based equivalents; weighted variants offer correlations similar to unweighted versions, but with a greater degree of spatial specificity; simplest-path methods confer an advantage in the context of local high-street mixed-uses but not necessarily for district-wide mixed-uses or land-use accessibilities; and the application of centrality measures to the dual network does not offer tangible benefits over the primal network.
\end{abstract}
\section{Street networks as emergent artefacts}\label{street-networks-emergent-artefacts}

Streets interconnect assortments of people and places. Whereas their purpose seems obvious, their emergent properties are less so. We intuitively sense that particular sidewalks are busier than others, that certain land-uses may have a proclivity towards distinct locations, or that highly successful streets can become public destinations in their own right. Although architects, urban designers, and planners may argue an instinctive appreciation for creating and nurturing such streets, history has often proved otherwise. Planned streets --- whether intentionally so or not --- have in many cases come to function antithetically to their historical purposes by prioritising connections to distant places at the expense of pedestrians and local connectivity. Streets and interstitial spaces can thus cease functioning as spatially connective tissue, leading to the implosion of public space while contributing to cities' social and functional fragmentation, a symptom of the broader paradigm of lower-density car-centric development characterised by a paucity of pedestrian-accessible land-uses. Nevertheless, even where the intention is to create more dense and walkable forms of urban development, it is not easy, without recourse to the incremental and evolutionary development typical of historical towns and cities, to anticipate the emergent characteristics of planned street networks and how these might relate to the potential vibrancy of land-uses.

Historical forms of urbanism evolved as an accretion of exchanges competing for access to the heart of villages, towns, and cities. An agglomerative dynamic, these processes feed --- and are fed by --- flows of people, goods, capital, and information cascading through the innumerable tendrils of urban networks reaching near or far. In congealed form, this logic is fine-grained and porous, with characteristically small urban blocks enmeshed by street networks interleaving numerous assortments of land-uses. \textcite{Jacobs1969} is quick to assure that any semblance of inefficiency is, actually, a hallmark of complex systems and the particular reason why evolutionary forms of urbanism can be persistently vibrant and resilient \parencite{Jacobs1961, Marshall2009}. Nevertheless, buoyed by the spirit of modernity, early 20th century architects, planners, and officials set out to tear these locally integrative networks asunder, mandating the rearrangement of cities around motor vehicles and the separation of land-uses \parencite{Hilberseimer1955, Garnier1989, Corbusier1967}. Though rationalised as ordered, efficient, and hygienic, the tree-like network structures deployed by modernists amounted to reductionism and a \emph{``compulsive desire for neatness and order''} \parencite[p.11]{Alexander1967}. By order of magnitude, these networks precluded the local combinatorial possibilities available on the densely interconnected semi-lattice structures of historical urban networks \parencite{Alexander1964}. A key ingredient had thus gone missing: planned street networks which prioritised motor vehicles seldom facilitated locally complex and evolving arrangements of interactions, the requisite lifeblood of healthy neighbourhoods and vibrant street-fronts \parencite{Harvey1989, Jacobs1961}.

The utopian cities envisioned by modernists were motivated by the zeitgeist of the times: belief in the power of rational thought and the invincibility of technological progress. New forms of mobility and communications technologies would compress space and time and (were supposed to) unify and connect the disparate communities and functionally segregated regions of planned cities. In reality, these networks marginalised pedestrians and, as epitomised by Robert Moses' heavy-handed slum-clearances in New York City, went so far as to lay waste to entire communities \parencite{Flint2011, Harvey1989, Lyon1999, Ellin1999}. The salient issue is that networks optimising space-time compression for certain citizens frequently create barriers for others. Problematically, an inordinate amount of power is wielded by those deciding or influencing the formation of such infrastructures and their consequent flows. These patterns represent an exclusionary model of urban development premised on the dissolution of public space, a state of urban disembodiment engendered through the network mechanisms of spatial bypass, selective access, and prohibitive pedestrian distances --- patterns we have come to associate with suburbia. \parencite{Graham2001, Pope1996}. These forms of planning not only remain prevalent but have found new life in the more recent Smart Cities hype \parencite{Greenfield2013, Hill2013, Townsend2013, Sterling2014}.

If planned street networks are brittle and biased or exhibit unpredictable emergent properties, then how should planned development proceed to preserve the cohesiveness of the public realm and enhance access for pedestrians? It is here where network analytic methods hold appeal. By allowing the opportunity to evaluate both existing and proposed street configurations, it becomes easier to make observations about their potential emergent properties as systems of streets. Thus personal intuition can be bridged to demonstrable, rigorous, and scalable forms of analysis while retaining integrity in light of complex systems interpretations of urbanism.
\section{The use of network analysis in urban design}\label{network-analysis-in-urban-design}

Early forms of network analysis arose within the context of the mathematical exploration of telecommunication networks \parencite{Shimbel1953} and have since gained widespread use and maturity in fields such as social network analysis \parencite{Wasserman1994}. The intuition is that networks (graphs) consisting of nodes (vertices) are connected by links (edges) to other nodes. For example, a social network may consist of people connected by relationships to other people. Network analysis can then be used to answer how connected or `central' a particular person is, thus inferring how important or influential they might be within the network's structure. Two of the more common measures of importance are \emph{closeness centrality} \parencite{Sabidussi1966}, how closely a node is located to other nodes, and \emph{betweenness centrality} \parencite{Freeman1977}, how often a node provides the link between other nodes. The computation of these measures involves the use of algorithms calculating \emph{shortest-paths} through the network: in the basic case, distance is topological --- the number of steps linking from node to node --- but it is also possible to represent distance by weighting the links between the nodes. For example, in street networks, these weights can correspond to the lengths of streets (links) between road intersections (nodes), and measures such as closeness and betweenness can then be used to infer how vital a particular location might be within the network.

To understand the broader context of street-network analysis in urban-design, it becomes necessary to distinguish between \emph{global} and \emph{local} forms of analysis; \emph{primal} and \emph{dual} street network representations; and \emph{shortest-path} as opposed to \emph{simplest-path} heuristics.

\begin{figure}[htbp]
 \centering
 \includegraphics[width=\textwidth, keepaspectratio]{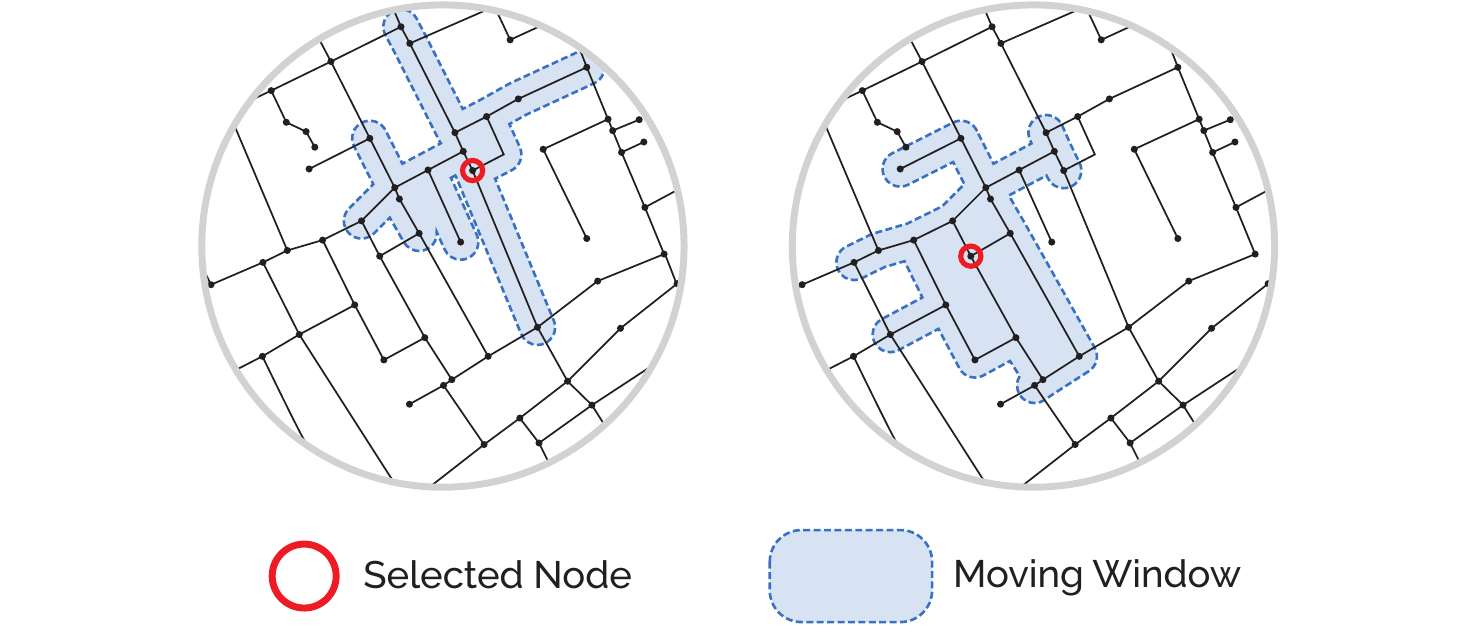}
 \caption[Moving-window methodology.]{For localised network analysis, metrics are calculated for each node in the network using a moving-window, which takes into account all other nodes within the specified distance threshold. In the simple case, these are based on crow-flies euclidean distances; in the more accurate case (shown), these are based on distances calculated over the street network.}\label{fig:moving_window_2}
\end{figure}

Network centrality methods can be applied to either the `global' or the `local' network. Global analysis scrutinises the entirety of the given network in contrast to localised forms of analysis which instead iteratively visit each node using a `windowing' methodology. This intuition is conveyed in Figure~\ref{fig:moving_window_2}: each node is visited, all reachable nodes within a specified distance are then identified and isolated from the network at large, and the analysis can then proceed on the isolated locally accessible network. The algorithm then steps to the next node and repeats the process until the entirety of the network is visited. Localised forms of analysis can be completed for a range of different distance thresholds, and these can be selected to concur with considerations such as pedestrian walking tolerances.

There are important implications for global as opposed to local methods: the computational outputs of global measures are directly coupled to the delineation of the overall network, which leads to the issue of how city boundaries are consistently defined and whether related metrics can be normalised in a sufficiently representative manner for accurate comparisons between different neighbourhoods, towns, and cities. A further drawback to global methods is that the assessment of different local planning configurations can be unnecessarily encumbered by distant regions of the network, which may have little impact on a pedestrian's perception of the walkability of local streets and neighbourhoods. From this point of view, whereas global properties of city networks can provide a wellspring of information about urban agglomerations and their evolutionary dynamics over time, they can also be abstruse and needlessly complex for the needs of urbanists and urban designers who are affected by more immediate and locally specific concerns: How walkable might a new development be? How many commercial or retail locations may be appropriate based on a 15-minute walk to a particular location? For these purposes, windowed network analysis methods hold more appeal because they remain contextually anchored; are capable of illustrating how properties of the pedestrian network are tangibly affected by day-to-day planning decisions and allow for the resultant measures to be consistently interpreted and compared from location to location or from city to city.

\begin{figure}[htbp]
 \centering
 \includegraphics[width=\textwidth, keepaspectratio]{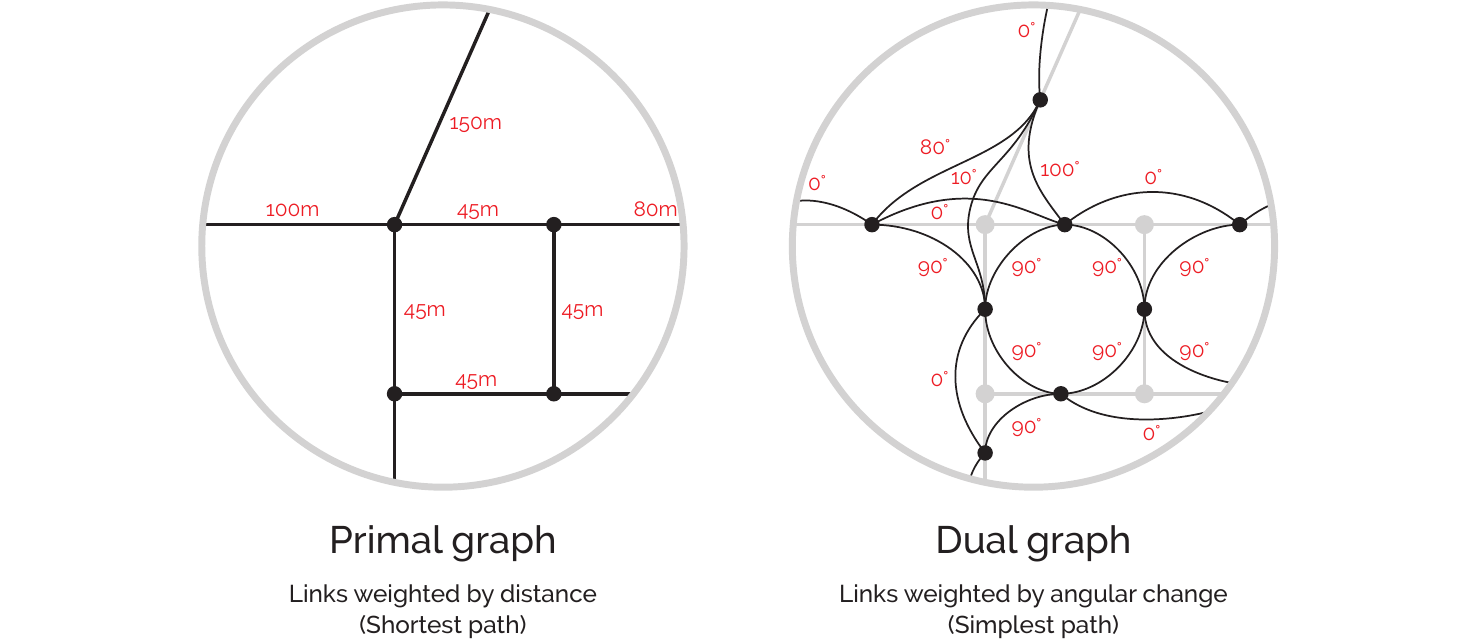}
 \caption[Primal and dual network representations.]{The primal street network representation (left) and a corresponding example of a dual representation (right). Note that different forms of dual representations exist: For example, traditional space syntax methods make use of axial lines that collapse adjacent straight segments into a single node (not shown).}\label{fig:primal_vs_dual}
\end{figure}

The \emph{primal} representation of street networks is when intersections are represented by nodes and streets by links, thus corresponding directly to conceptions of streets embedded in euclidean space: intersections adopt specified coordinates connected by streets to other intersections. However, it is also possible to adopt the \emph{dual} representation where, inversely, nodes correspond to streets and links correspond to the connections between them \parencite{Rosvall2005, Marshall2018}. In simpler cases, per Figure~\ref{fig:primal_vs_dual}, the dual network can be a direct inversion of the primal. However, there are also other more complex procedures for generating dual-network topologies, such as using the angular straightness of streets or considerations such as street names as a basis for `collapsing' adjacent street segments into a single (dual representation) node which may then be individually linked to a potentially large number of other adjacent (dual representation) nodes. These types of procedure can be used to extricate the dual representation from its euclidean embedding, thus sacrificing spatial proximities in favour of a more abstract informational representation consisting of topological links \parencite{Batty2004}. Such approaches can be used to analyse properties such as the global connectivity of the network and its \emph{small-world} and \emph{scale-free} characteristics, which are emblematic of complex systems and network analysis more generally \parencite{Albert2002, Porta2006a}.

\begin{figure}[htbp]
 \centering
 \includegraphics[width=\textwidth, keepaspectratio]{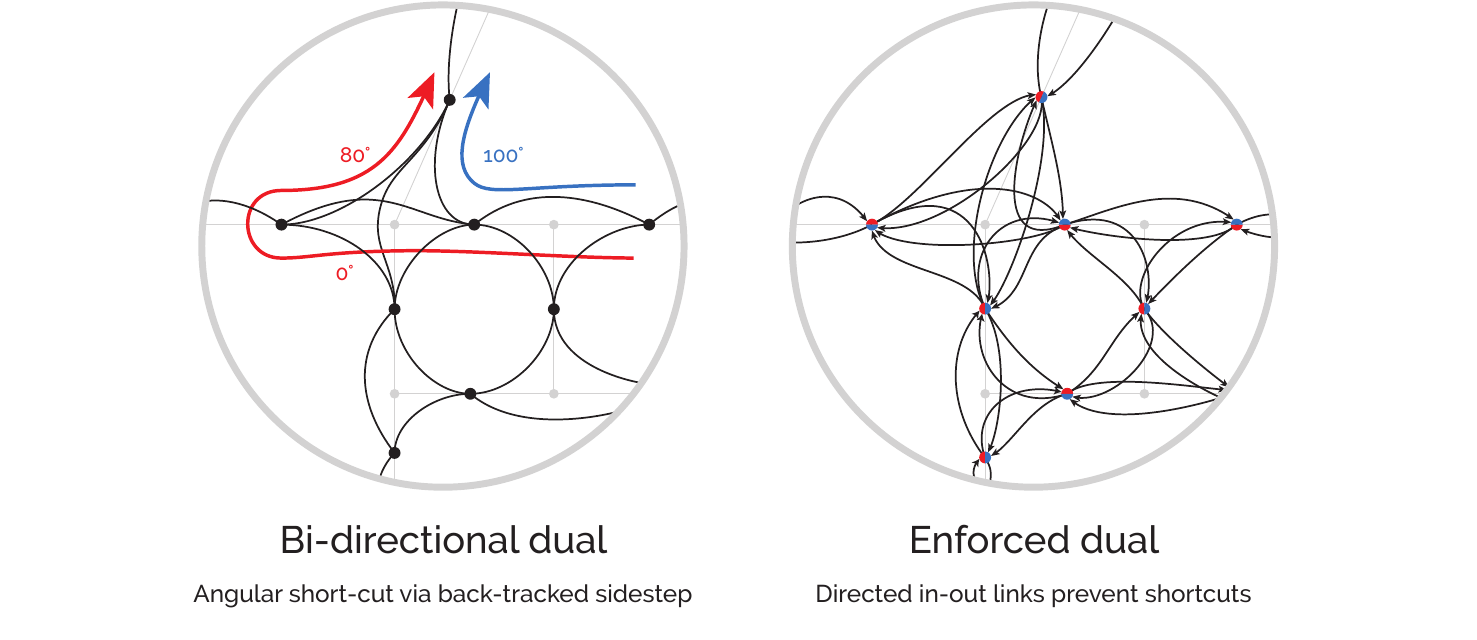}
 \caption[Shortest-path sidestep-problem on angular weighted networks.]{On dual networks using angular distances, shortest path algorithms can sidestep sharp angles by combining adjacent shallower angles (left). This necessitates enforced in and out directions for each node (right) to ensure that the algorithm continues-on in the same direction as which it arrived.}\label{fig:enforced_dual}
\end{figure}

A prevalent form of urban network analysis --- \emph{space syntax} \parencite{Hillier1984} --- emerged around the use of the dual representation in order to quantify changes in direction from one street to another. Euclidean distances are thereby waived in favour of the number of topological turns or, per more recent interpretations, the cumulative angular changes in direction from turn to turn (see Figure~\ref{fig:primal_vs_dual}). Straighter routes with better lines-of-sight are, therefore, considered `shorter' than more convoluted routes even if the euclidean distances are greater, thus transforming the emphasis from \emph{shortest} paths to \emph{simplest} paths. This strategy reflects the space syntax premise that cognitive way-finding in relation to lines-of-sight take precedence to the distances traversed between locations and that routes requiring minimal geometric complexity are favoured to routes requiring minimal physical effort \parencite{Hillier2005, Serra2019}; accordingly, geometric and topological properties of the street network are deemed primary determinants in the general evolution of land-uses and the wider patterns of activities in cities \parencite{Penn1998, Hillier2007}. Space syntax concepts, history, and terminology is similar but not necessarily the same as that used more generally by the network-analysis scientific community. This can lead to some ambiguity on points such as terminology, e.g.~\emph{mean depth} instead of \emph{normalised farness}, \emph{integration} instead of \emph{normalised closeness}, or \emph{choice} instead of \emph{betweenness}. Some of the methods and interpretations used by space syntax may consequently appear esoteric to outsiders: by way of example, where techniques on the extrapolation of \emph{axial lines} (uninterrupted lines of sight) from the street network led to drawn-out deliberations on how exactly this should be done and whether it is possible to do so in an algorithmically rigorous manner \parencite{Turner2005, Porta2006, Ratti2004}. Newer and more tractable methods have alleviated this particular debate --- \emph{fractional analysis} and the now prevalent \emph{angular segment analysis} --- compute total angular change along road centrelines instead of counting topological turns from axial line to axial line \parencite{Turner2000, Dalton2001, Turner2005a, Turner2007}. Nevertheless, this has introduced a new algorithmic challenge in that simplest-path algorithms will `side-step' sharp angular turns in cases where smaller combinations of adjacent angles can be combined instead (Figure~\ref{fig:enforced_dual}), thus making it necessary to enforce in and out directions for algorithms as they pass-through nodes during graph traversals.

Simplest vs shortest-path heuristics should not be construed as an either-or situation because the inter-relation is varied \parencite{Barthelemy2015} and assertions to the contrary should be treated with caution. Pedestrian route choices are complex and can be based on a combination of shortest distances and cognitively simplest routes, and it can be reasoned that this may vary from place to place and from person to person. Whereas simplest-path centralities do not directly take distances into account, and shortest-path centralities do not directly take angular deviation into account, the reason why either of these may work in isolation is that shortest routes are frequently the same as the simplest \parencite{Viana2013, Omer2018}. Further, strong correlations are observed between shortest and simplest path heuristics at certain distances \parencite{Hillier2007, Serra2019}. Where multiple compelling route options exist, the mechanisms behind route choices are far from clear-cut. For example, a tourist might rely on lines-of-sight to explore a city via boulevards and significant landmarks, but may, in other cases, follow shortest-paths through the use of a map or a mobile device. Similarly, a resident might rely on sight to find a pub or deli in an unfamiliar neighbourhood but may otherwise have access to an assortment of locally familiar locations, potentially hidden around corners off-the-beaten-path to be fluidly connected in manifold non-simple route permutations influenced by varied daily considerations. Modal choice presents an additional nuance, wherein it could be argued that pedestrians --- in contrast to motor vehicles --- have a lower resistance to turns but a higher resistance to farther distances and may thus actively choose routes that are shorter or less simple than the more axial routes frequented by vehicles \parencite{Serra2019}. In yet other cases, pedestrians may opt for routes that are neither the shortest nor the simplest but offer a more pleasant alternative free from traffic, noise, or fumes; or greater access to greenery, sunlight, shade, or shallower gradients. In short, pedestrian route-choice considerations are varied, and simplest or shortest-path methods can be thought of as proxies to some of the more prevalent patterns of spatial flows.

Research has shown support for \emph{simplest} path measures within the context of vehicular travel, particularly at distances larger than 2000m. However, there are hints that \emph{shortest} path measures retain relevancy for smaller distances and non-vehicular forms of transport such as cycling \parencite{Serra2019}, thus prompting broader questions pertinent to urbanists and urban designers: specifically, which methods work well at the pedestrian scale regarding the intensity and mix of walkable land-uses?

Whereas shortest-path routes are ordinarily applied to primal representations and simplest-path routes are most often applied to dual representations, these concepts should not necessarily be conflated. The primal representation can be used with simplest-path angular-segment-analysis methods if the algorithms are designed accordingly; likewise, dual representations can be used with shortest-path methods if the dual network has not been removed from its spatial embedding. There may, further, be potential synergies to be gained through the simultaneous application of both primal and dual representations and the respective methods of inquiry suited to either approach \parencite{Masucci2016}.
\section{Technical considerations for urban centrality methods}\label{the-measures}

\subsection{Local distance thresholds: crow-flies vs.~network-distances}

\begin{figure}[ht]
 \centering
 \includegraphics[width=\textwidth, keepaspectratio]{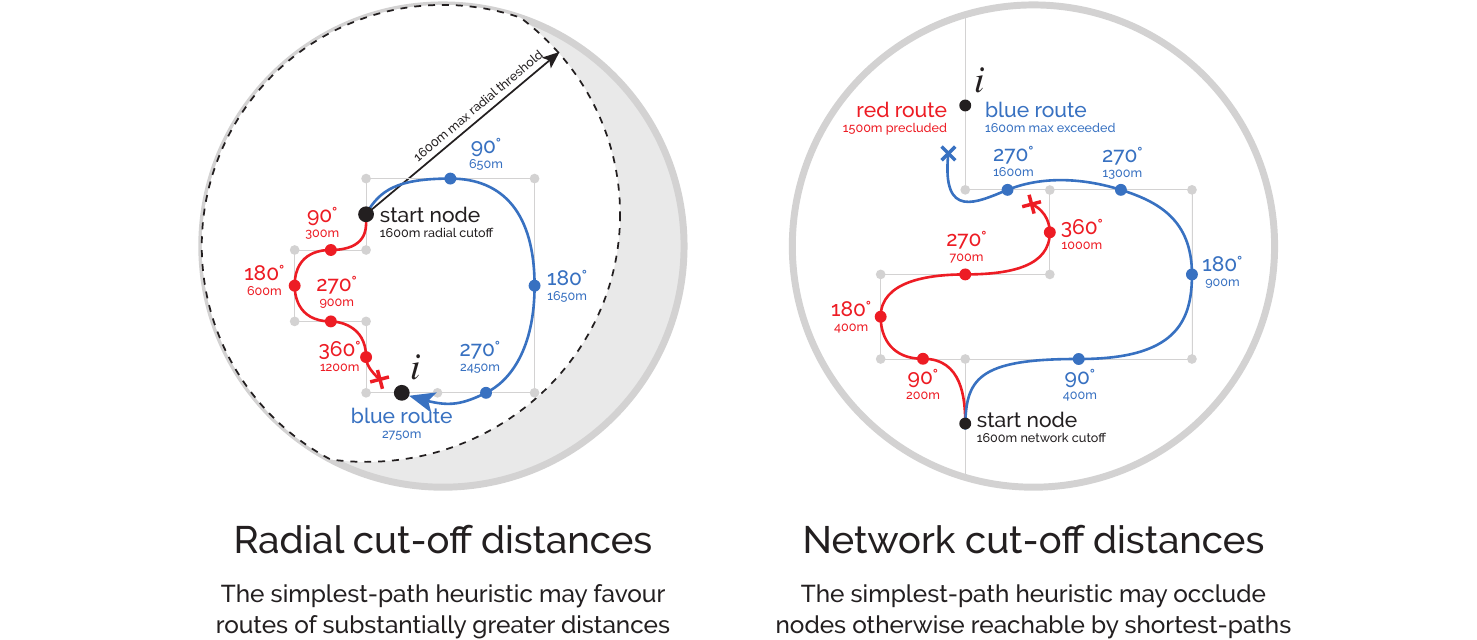}
 \caption[Implications of shortest and simplest path heuristics.]{Crude radial distance thresholds (left) can permit distances significantly exceeding the threshold, in some cases also overshooting the threshold boundary. Conversely, the use of strict network distance thresholds (right) will in some cases occlude nodes that may otherwise have been within the maximum distance threshold if using shortest-paths.}\label{fig:simplest_path_heuristic}
\end{figure}

Localised measures are similar to the global equivalents from which they derive but only take nodes up to selected distance thresholds into account. The procedure entails a moving-window approach (Figure~\ref{fig:moving_window_2}), wherein the algorithm iteratively visits each node from which it then proceeds to isolate all other nodes within the maximum distance threshold. A contradiction is introduced for simplest-path analysis \parencite{Cooper2015}: the euclidean distance metric used for the threshold differs from the angular distance metric used for the analysis, thus introducing an implicit euclidean component. Further, when using simple radial (as opposed to network) distance cutoffs, this combination can lead simplest-path algorithms to behave in unexpected ways, such as selecting routes traversing network distances that substantially exceed the implied distance of a given radial cutoff: in some cases, even overshooting the threshold then turning-back to end inside the radial boundary. Various strategies can be considered to mitigate such behaviour, and the position taken in this analysis is that --- in keeping with the implied distance of a stated threshold --- the distances traversed by an algorithm should be strictly enforced based on actual network distances. Therefore, intuition and behaviour remain consistent regardless of the heuristic and correspond more realistically to available routes for a given pedestrian walking tolerance. In some cases, nodes that may have been within maximum network distance thresholds as measured by shortest-paths will be occluded by the sometimes greater distances encountered along simplest-paths, which is here accepted as an inherent and representative property of the routes preferred by the simplest-path heuristic (Figure~\ref{fig:simplest_path_heuristic}).

\subsection{Topological structures and related distortions}

Network analysis algorithms are, of necessity, sensitive to the topological structure of the network, and poor quality network datasets will lead to spurious results. For example, broken links will throw shortest-path algorithms off course, and unnecessarily complex representations of features such as complex road intersections, private roadways in housing estates, or individually mapped aisles in large parking lots will introduce incorrect centralities to the wider network. This situation arises because network analysis methods are susceptible to topological distortions arising from variations in the intensity of nodes for equivalent lengths of road segments. For example, a straight street segment may be represented as a single link between two nodes, whereas an arced street of equivalent length is often represented with additional nodes to approximate the curvature of the roadway. Each additional node results in more summations for all other nodes within windowed distance thresholds, and this effect is further magnified when such nodes are densely co-located, consequently skewing the outcome of centrality measures. The \emph{Ordnance Survey} \emph{Open Roads} dataset used in the following analysis is of particularly high quality: it is already cleaned and simplified. It keeps the geometric representation of roadways distinct from the topological structure of the network, thus minimising the stated issues. If using topologically `messier' networks from a source such as \href{https://www.openstreetmap.org}{Open Street Map} then it is generally necessary to first clean and simplify the network, which should be done while preserving the geometry of the original street segments so that accurate distances or angular changes in direction can be recovered. These techniques are facilitated by the \code{cityseer-api} package and are discussed more comprehensively in
\ifpaper \textcite{Simons2021b}.
\else Chapter~\ref{the-cityseer-api-Python-package}.
\fi
Note that automated cleaning routines which merge nodes and street geometries can result in some geometrical `quirks' in the form of small twists or turns. These tend to be sufficiently minor that shortest-path methods are not substantially degraded; however, it may be necessary to manually check for downstream artefacts affecting simplest-path calculations (when using simplest-path methods). 

\begin{figure}[htp]
 \centering
 \includegraphics[width=\textwidth, keepaspectratio]{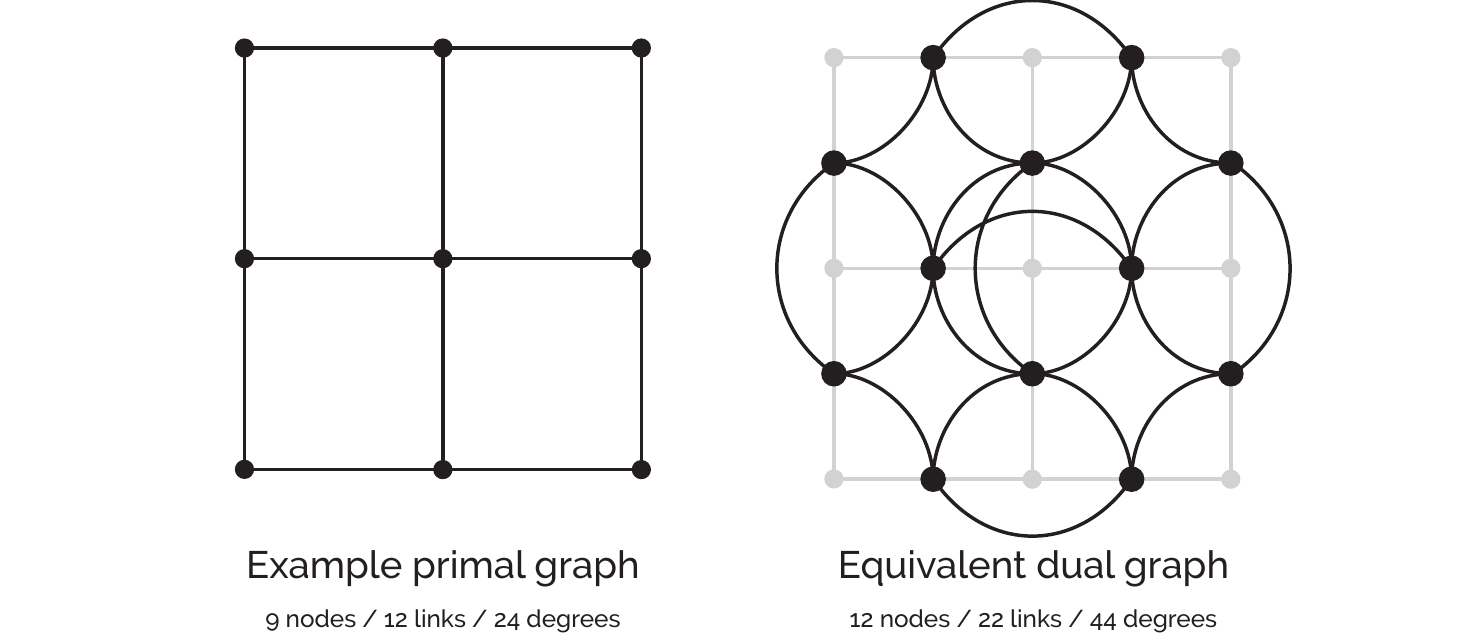}
 \caption[Topological variances between primal and dual network representations.]{Even if the underlying structure is the same, primal and dual representations are not directly comparable because they have distinct topological characteristics. Measures computed on the dual network will demonstrate higher effective centralities due to greater quantities of nodes and links, and this effect is amplified for densely interconnected networks.}\label{fig:primal_vs_dual_structure}
\end{figure}

Another manifestation of this phenomenon is the topological divergence between primal and dual representations (see Figure~\ref{fig:primal_vs_dual_structure}). Even though both derive from the same underlying structure, the outputs for the same network measures using the same weights will yield different results; the dual representation generates higher equivalent metrics due to larger quantities of nodes and links, which becomes more pronounced as the complexity of the network increases. Comparisons between shortest-path measures on primal networks and simplest-path measures on dual networks must therefore be avoided. The \code{cityseer-api} package computes angular weights using a dynamic methodology that accumulates changes from road segment to road segment, thereby allowing simplest-path methods to be computed on the primal network and facilitating the direct comparison of shortest and simplest-path (angular-segment-analysis) methods on the same topological networks.

Note that whereas it may be tempting to resolve poor-quality networks or topological divergences through normalisation by the number of nodes, this would prevent accurate comparisons from location to location (discussed at length below). Another strategy is to weight the nodes by factors such as street lengths \parencite{Turner2007} or the number of adjacent buildings \parencite{Sevtsuk2012}. The following analysis considers two such strategies: use of segmented (continuous) forms of `length weighted' centralities and the optional use of network decomposition to distribute nodes more evenly across networks as an implicit form of length normalisation, with the benefit that variations can then be captured at a higher resolution of analysis within road segments
\ifpaper \parencite{Simons2021b}.
\else (see Section~\ref{the-cityseer-api-Python-package}).
\fi

\subsection{Closeness centrality}

The \emph{closeness} measure 
\begin{equation}\label{eq:closeness}
 Closeness_{(i)} = \frac{1}{\sum_{j\neq{i}}d_{(i,j)}}
\end{equation}
is the reciprocal of \emph{farness}
\begin{equation}\label{eq:farness}
 Farness_{(i)} = \sum_{j\neq{i}}d_{(i,j)}\, ,
\end{equation}
where farness is the sum of distances $d$ from node $i$ to all reachable nodes $j$ \parencite{Sabidussi1966}. Closeness conveys how proximate node $i$ is to surrounding nodes, with the implication that high closeness centralities afford increased likelihood of access and interaction. Closeness can be normalised by the number of nodes in the graph $N$ (not including node $i$, which gives $N-1$) divided by the sum of the distances
\begin{equation}\label{eq:normalised_closeness}
 Normalised\ C_{(i)} = \frac{N-1}{\sum_{j\neq{i}}d_{(i,j)}}\, ,
\end{equation}
which is the inverse of the arithmetic mean (average) of farness
\begin{equation}\label{eq:normalised_farness}
 Normalised\ F_{(i)} = \frac{\sum_{j\neq{i}}d_{(i,j)}}{N-1}\, .
\end{equation}

\begin{figure}[htp]
 \centering
 \includegraphics[width=\textwidth, keepaspectratio]{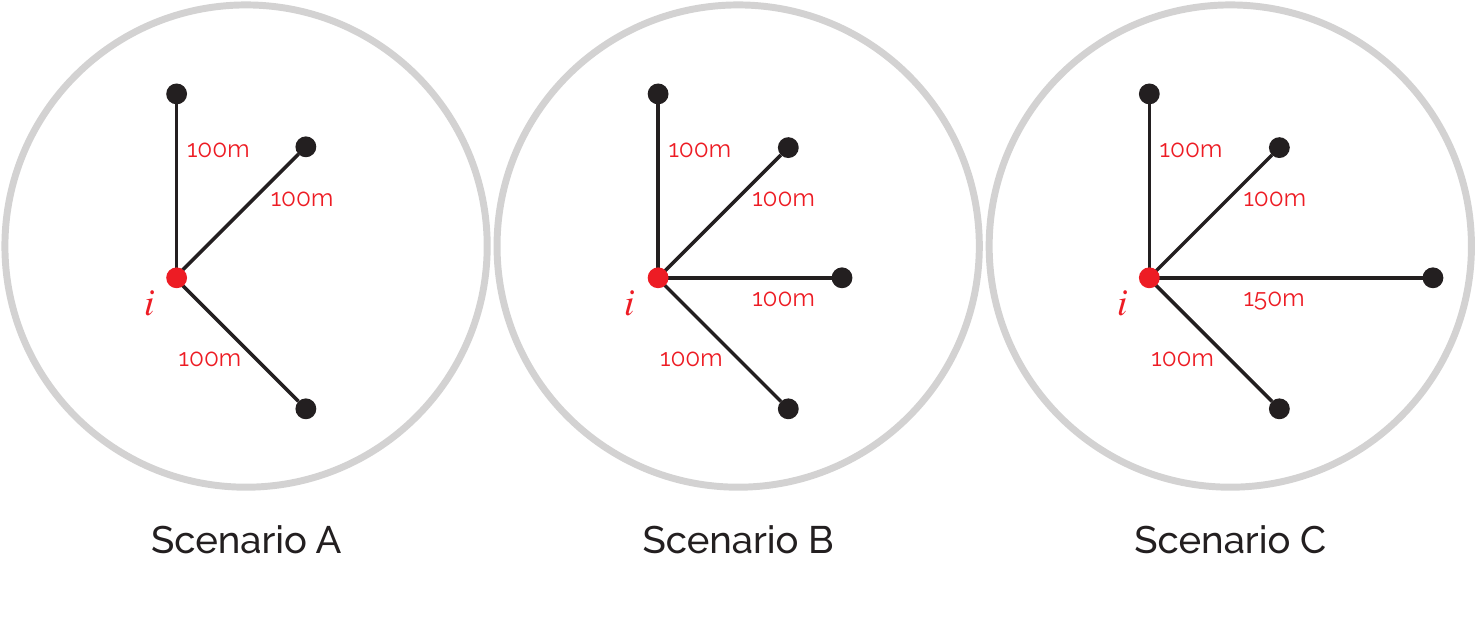}
 \caption{Simple comparative localised closeness scenarios.}\label{fig:closeness_comparisons}
\end{figure}

The normalised form of Closeness tends to be the most commonly applied derivation in urban analysis; however, closer inspection reveals that a series of challenges arise when applied to localised (windowed) network analysis. Per Figure~\ref{fig:closeness_comparisons} and Table~\ref{table:closeness_comparisons}, normalised closeness does not behave as intended for disconnected or `windowed' graphs, in which cases non-normalised harmonic closeness centrality \parencite{Marchiori2000, Rochat2009} of the form
\begin{equation}\label{eq:harmonic_closeness}
 Harmonic\ C_{(i)} = \sum_{j\neq{i}}\frac{1}{d_{(i,j)}}
\end{equation}
behaves more suitably, with the difference being that the division happens prior to the summation. For context, the normalised form of harmonic closeness centrality is the inverse of the harmonic mean of farness (as opposed to the arithmetic mean in~\ref{eq:normalised_farness})
\begin{equation}\label{eq:normalised_harmonic_closeness}
 Normalised\ HC_{(i)} = \frac{\sum_{j\neq{i}}\frac{1}{d_{(i,j)}}}{N-1}\, .
\end{equation}

Another option is to use \emph{improved closeness} centrality proposed by \textcite{Wasserman1994}
\begin{equation}\label{eq:improved_closeness}
 Improved\ C_{(i)} = \frac{(N-1)_{i}/{(g-1)}}{\sum_{j\neq{i}}d_{(i,j)}/{(N-1)_{i}}}\, ,
\end{equation}
which is intended for situations where a limited subset of nodes is reachable, and is defined as the ratio of the fraction of reachable nodes $(N-1)_{i}$ to the average distance to those nodes. The global number of nodes $g$ is unknown, but since this is effectively the worldwide street network shared by all subgraphs it can be argued that the reachable nodes $(N-1)_{i}$ can forego the normalisation by $(g-1)$, thus giving:
\begin{equation}\label{eq:used_closeness}
 C_{(i)} = \frac{(N-1)_{i}}{\sum_{j\neq{i}}d_{(i,j)}/{(N-1)_{i}}} = \frac{(N-1)_{i}^2}{\sum_{j\neq{i}}d_{(i,j)}}\, .
\end{equation}
Note that if global nodes $g$ were instead defined as the `parent' city, then results would not be comparable between locations, and this would otherwise become problematic when arbitrary boundaries delineate networks spanning adjacent towns or cities. The intuition of the resultant form can be contrasted to normalised Closeness (Eq:~\ref{eq:normalised_closeness}) with the number of nodes $(N-1)_{i}$ being divided by the average instead of the total sum of distances to those nodes. Improved Closeness, therefore, behaves more intuitively: it increases either for a greater number of locally accessible nodes or if the average distance to those nodes decreases.

Figure~\ref{fig:closeness_comparisons} illustrates these formulations more intuitively. Scenario B should be expected to have a greater closeness within an urban context than A because it is equivalently close to a greater number of nodes. Scenario C contains a more distant node and should therefore have a lower closeness centrality than Scenario B but should still be greater than A.

\begin{figure}[htp]
 \centering
 \includegraphics[width=\textwidth, keepaspectratio]{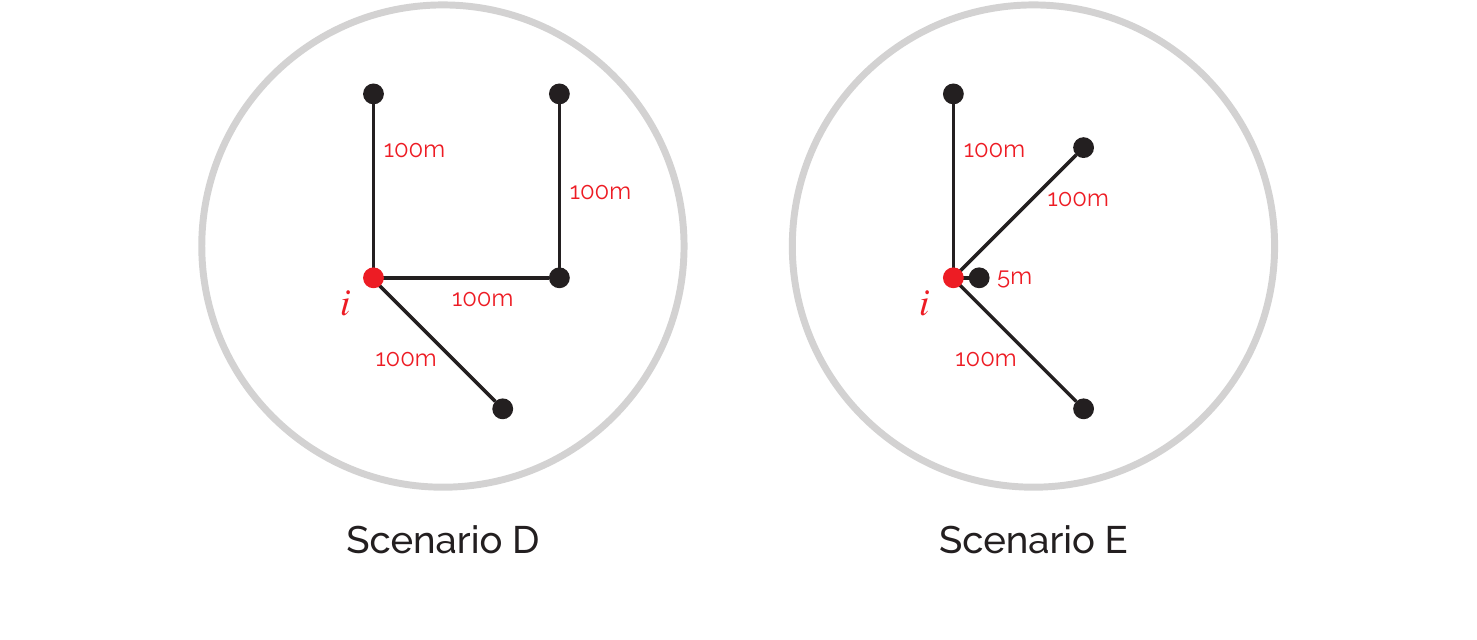}
 \caption{Localised closeness scenarios with ``very nearby'' nodes.}\label{fig:closeness_comparisons_near_nodes}
\end{figure}

As shown in Table~\ref{table:closeness_comparisons}, Closeness counter-intuitively decreases from Scenario 1 to 2, whereas normalised Closeness is also problematic because it neutralises variations between differently sized windowed subgraphs, rendering them incomparable. Harmonic closeness centrality and the modified form of improved closeness centrality variants behave as anticipated. Out of these, harmonic Closeness is technically more precise because it considers the inverse of the distances independently in contrast to improved Closeness, which first averages the distances; yet, for the same reason, the averaging implicit with improved Closeness may be advantageous when working with poorer quality or un-simplified representations of street networks. This is because nodes placed particularly close to other nodes would otherwise have an unexpectedly large impact on the summed closeness (see Figure~\ref{fig:closeness_comparisons_near_nodes} and Table~\ref{table:closeness_comparisons_near_nodes}). The issue is exacerbated by simplest-path centralities because there are situations where angular distances approach zero for small local distance thresholds, even when averaged, thus causing infinitely large values through division by zero.

\begin{table}[htp]
 \centering\footnotesize
 \begin{tabular}{ r | r r r }
 &
 Scenario A &
 Scenario B &
 Scenario C \\
 \midrule
 $Farness_{(i)}$ &
 $100+100+100=300$ &
 $100+100+100+100=400$ &
 $100+100+150+100=450$ \\

 $Norm.\ F_{(i)}$ &
 $\frac{100+100+100}{3}=100$ &
 $\frac{100+100+100+100}{4}=100$ &
 $\frac{100+100+150+100}{4}=112.5$ \\

 $Closeness_{(i)}$ &
 $\frac{1}{100+100+100}=0.00\overline{3}$ &
 $\frac{1}{100+100+100+100}=0.0025$ &
 $\frac{1}{100+100+150+100}=0.00\overline{2}$ \\

 $Norm.\ C_{(i)}$ &
 $\frac{3}{100+100+100}=0.01$ &
 $\frac{4}{100+100+100+100}=0.01$ &
 $\frac{4}{100+100+150+100}=0.00\overline{8}$ \\

 $Harmonic\ C_{(i)}$ &
 $\frac{1}{100} + \frac{1}{100} + \frac{1}{100}=0.03$ &
 $\frac{1}{100} + \frac{1}{100} + \frac{1}{100} + \frac{1}{100}=0.04$ &
 $\frac{1}{100} + \frac{1}{100} + \frac{1}{150} + \frac{1}{100}=0.03\overline{6}$ \\

 $Norm.\ HC_{(i)}$ &
 $(\frac{1}{100} + \frac{1}{100} + \frac{1}{100}) / 3=0.01$ &
 $(\frac{1}{100} + \frac{1}{100} + \frac{1}{100} + \frac{1}{100}) / 4=0.01$ &
 $(\frac{1}{100} + \frac{1}{100} + \frac{1}{150} + \frac{1}{100}) / 4=0.0092$ \\

 $Improved\ C_{(i)}$ &
 $\frac{3}{(100+100+100)/3}=0.03$ &
 $\frac{4}{(100+100+100+100)/4}=0.04$ &
 $\frac{4}{(100+100+150+100)/4}=0.03\overline{5}$ \\
 \end{tabular}
 \caption{Closeness comparisons}\label{table:closeness_comparisons}
\end{table}

\begin{table}[htp]
 \centering\footnotesize
 \begin{tabular}{ r | r r r }
 &
 Scenario D &
 Scenario E \\
 \midrule
 $Harmonic\ C_{(i)}$ &
 $\frac{1}{100} + \frac{1}{100} + \frac{1}{100} + \frac{1}{200} =0.035$ &
 $\frac{1}{100} + \frac{1}{100} + \frac{1}{5} + \frac{1}{100} =0.23$ \\
 $Improved\ C_{(i)}$ &
 $\frac{4}{(100+100+100+200) / 4}=0.032$ &
 $\frac{4}{(100+100+5+100) / 4}=0.052$ \\
 \end{tabular}
 \caption[Comparative closeness scenarios showing indirect segments and very short segments.]{Comparative closeness scenarios showing indirect segments and very short segments. Improved Closeness is less susceptible to large numeric variations introduced by very short street segments.}\label{table:closeness_comparisons_near_nodes}
\end{table}

The effects of particularly nearby nodes segue into a broader issue: topological distortions due to varying intensities of nodes can distort the outcome of centrality measures. For this reason, it can be beneficial to consider the measures in a street-length-weighted form, particularly if dealing with `messy' sources of graph topology. In this form, a node's contribution to centrality is weighted by the length of the adjacent street segments because greater exposure to street lengths offers a greater potential for interaction. Accordingly, the contribution of unusually high concentrations of nodes would be tempered by correspondingly shorter street segments and vice-versa. In practical terms, half of the length of a street segment is assigned to the node on either end, with the aggregated lengths per street node $l$ then used to weight the measure in the numerator:
\begin{equation}\label{eq:harmonic_length_weighted}
 length\ wt.\ HC_{(i)} = \sum_{j\neq{i}}\frac{l}{d_{(i,j)}}\, .
\end{equation}
Although the approach seems relatively straightforward, a drawback is that continuously decaying spatial impedances (e.g.~$1/d$) are first discretised at node locations and are then, in effect, mapped back onto the adjacent street segments through the process of street-length weighting, resulting in step-wise (rather than continuous) increases in spatial impedances as distances increase over the network. A further implication is that the half-segments adjacent to the origin node must be ignored; otherwise, infinitely large Closeness arises through division by zero distance.

An alternative is to use the integral of harmonic closeness. The integral for $f(x)=1/d$ takes the form
\begin{equation}
 \int_{a}^{b}\ f(x)\ dx = \ln(b) - \ln(a)
\end{equation}
and can be applied to sum the `area under the (spatial impedance) curve' for the respective lower and upper segment bounds $a$ and $b$ for all reachable segments $S$:
\begin{equation}\label{eq:harmonic_integral}
 continuous\ HC_{(i)} = \sum_{(a, b)}^{S} \int_{a}^{b}\ f(x)\ dx = \sum_{(a, b)}^{S}\ \ln(b) - \ln(a)\, .
\end{equation}
This allows spatial impedances to increase continuously. For example, a 10m street segment adjacent to the origin is now found as $\ln(10) = 2.303$ and a segment from $10m$ to $20m$ distant is found as $\ln(20) - \ln(10) = 0.693$. The continuous form means that aggregations remain consistent regardless of how many times street lengths are split at intervening nodes, i.e.~$\ln(20) = \ln(10) + (\ln(20) - \ln(10)) = 2.996$, and this form is therefore relatively robust if applied to `messy' graph topologies.

\begin{figure}[htp]
 \centering
 \includegraphics[width=\textwidth, keepaspectratio]{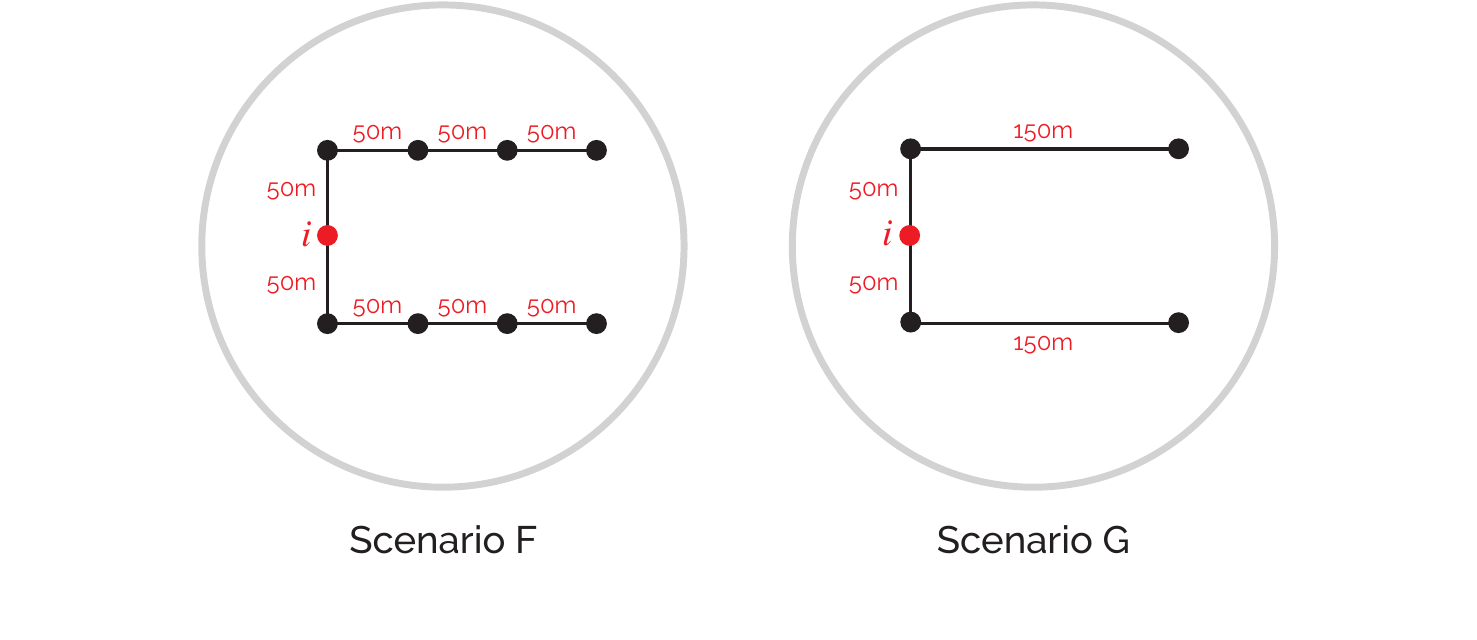}
 \caption{Localised closeness scenarios comparing varied node intensities.}\label{fig:closeness_comparisons_length_weighted}
\end{figure}

\begin{table}[htp]
 \centering\footnotesize
 \begin{tabular}{ r | r r }
 &
 $length\ wt.\ HC_{(i)} = \sum_{j\neq{i}}\frac{l}{d_{(i,j)}}$ &
 $Harmonic\ C_{(i)} (0, x) = \int_{0}^{x} \ln(x)\ dx$ \\
 \midrule
 Scenario A &
 $\frac{50}{100} + \frac{50}{100} + \frac{50}{100} = 1.5$ &
 $\ln(100) + \ln(100) + \ln(100) = 13.82$ \\
 Scenario B &
 $\frac{50}{100} + \frac{50}{100} + \frac{50}{100} + \frac{50}{100} = 2.0$ &
 $\ln(100) + \ln(100) + \ln(100) + \ln(100) = 18.42$ \\
 Scenario C &
 $\frac{50}{100} + \frac{50}{100} + \frac{75}{150} + \frac{50}{100} = 2.0$ &
 $\ln(100) + \ln(100) + \ln(150) + \ln(100) = 18.83$ \\
 Scenario D &
 $\frac{50}{100} + \frac{100}{100} + \frac{50}{100} + \frac{50}{200} = 2.25$ &
 $\ln(100) + \ln(100) + \ln(100) + (\ln(200) - \ln(100)) = 14.51$ \\
 Scenario E &
 $\frac{50}{100} + \frac{50}{100} + \frac{2.5}{50} + \frac{50}{100} = 2$ &
 $\ln(100) + \ln(100) + \ln(100) + \ln(5)= 15.42$ \\
 Scenario F &
 $(\frac{50}{50} + \frac{50}{100} + \frac{50}{150} + \frac{25}{200})*2 = 3.92$ &
 $(simplified)\ \ln(200) * 2 = 10.60$ \\
 Scenario G &
 $(\frac{100}{50} + \frac{75}{200})*2 = 4.75$ &
 $(simplified)\  \ln(200) * 2 = 10.60$ \\
 \end{tabular}
 \caption[Length weighted closeness comparisons]{Length weighted closeness comparisons comparing street-length weighted harmonic closeness and a continuous form of harmonic closeness.}\label{table:closeness_comparisons_length_weighted}
\end{table}

The length-weighted (\ref{eq:harmonic_length_weighted}) and continuous form (\ref{eq:harmonic_integral}) of harmonic closeness are compared in Table~\ref{table:closeness_comparisons_length_weighted} across all above scenarios, including two additional scenarios detailed in Figure~\ref{fig:closeness_comparisons_length_weighted}. In length-weighted harmonic Closeness, the street segments are divided and assigned to the respective nodes on either side, meaning that the portions adjacent to the origin nodes are not considered. The measure generally performs as intended, but the lack of accommodation for origin segments presents some difficulties; for example, Scenario C yields an equivalent value to Scenario B, and Scenario D counter-intuitively yields a higher value than B or C. Further, differences in node densities per F and G are not neutralised despite weighting by segment lengths. The continuous form resolves the issues mentioned above and behaves intuitively: Scenario D is greater than A but less than B or C, Scenario E is greater than A but less than B, Scenario C is greater than B, and different node densities are neutralised per F and G.

Additional nuances need consideration in the case of angular centralities. Whereas forms of harmonic or improved Closeness are suitable, a decision must be made about how angular distances are scaled. \textcite{Turner2007, Hillier2005} have suggested scaling to 0 for no turns and 2 for 180\degree turns; however, in the case of harmonic Closeness, this would leave the division through zero conundrum unresolved and would thus imply that pedestrians would walk infinitely far on streets with 0\degree angular impedance. A scaling of 1 through 3 may be preferable to cap a node's contribution to the unit length $l$ instead of $\infty$. In the case of street-length weighted variants, the continuous form suggested per Equation~\ref{eq:harmonic_integral} cannot be used because simplest-path impedances do not increase continuously. An alternative strategy can be considered wherein segment lengths in metres are divided by angular impedances, e.g. a $100m$ segment encountered at $100m$ distance from the origin at 0\degree of angular change would contribute $\frac{100}{1}$ whereas the equivalent segment encountered around a 90\degree corner would be divided by 2; by 3 in the case of 180\degree; etc.

Note that segment-based methods should not be computed on the dual graph because the resultant proliferation of overlapping edges would result in duplicitous summations compared to the originating primal street edges.

\subsection{The gravity index}

The use of spatial impedance as an accessibility measure, which is often referred to as the \emph{gravity index}, offers a more elegant approach to the weighting of distances. In contrast to street network analysis, which emphasises the physicalism of the road network structure as an artefactual manifestation of flows, `gravity' based methods based on impedances shift the emphasis to the potential flow of interactions as they might unfold through space from the nodes across such structures \parencite{Batty2013, Hansen1959, Rutherford1979}. The distinction may seem pedantic, and whereas the spatial impedance function bears some resemblance to harmonic closeness centrality, it introduces a more sophisticated approach in that it more explicitly models spatial impedances. It accordingly reflects the potential for spatial interaction from nodes $j$ to node $i$ and --- like gravity --- how the potential for this interaction decays with distance. This typically takes the form of the negative exponential
\begin{equation}\label{eq:gravity}
 Gravity_{(i)}=\sum_{j\neq{i}}\exp(-\beta\cdot d)
\end{equation}
where the rate of decay $\beta$ (in the negative exponential) can be set to model specific trip-purposes or transportation modes to reflect people's willingness to travel a given distance $d$ to particular types of locations \parencite{Sevtsuk2012, Handy1997, Iacono2008}.

Note that the term \emph{gravity index} can be somewhat misleading: full-fledged gravity models form the basis of broader land-use and transportation modelling where the attraction of the origins and destinations are, as per gravity, taken into account. However, the \emph{gravity index} generally assumes equal attractions from each node to every other node and, as such, is simply a means to provide distance-weighted counts of accessible locations $j$ proximate to $i$ at the given impedance $\beta$. This approach works well for quantifying access to specific land-uses or, in the case of the street network structure, quantifying physical access from nodes $j$ to node $i$.

\begin{figure}[htp]
 \centering
 \includegraphics[width=\textwidth, keepaspectratio]{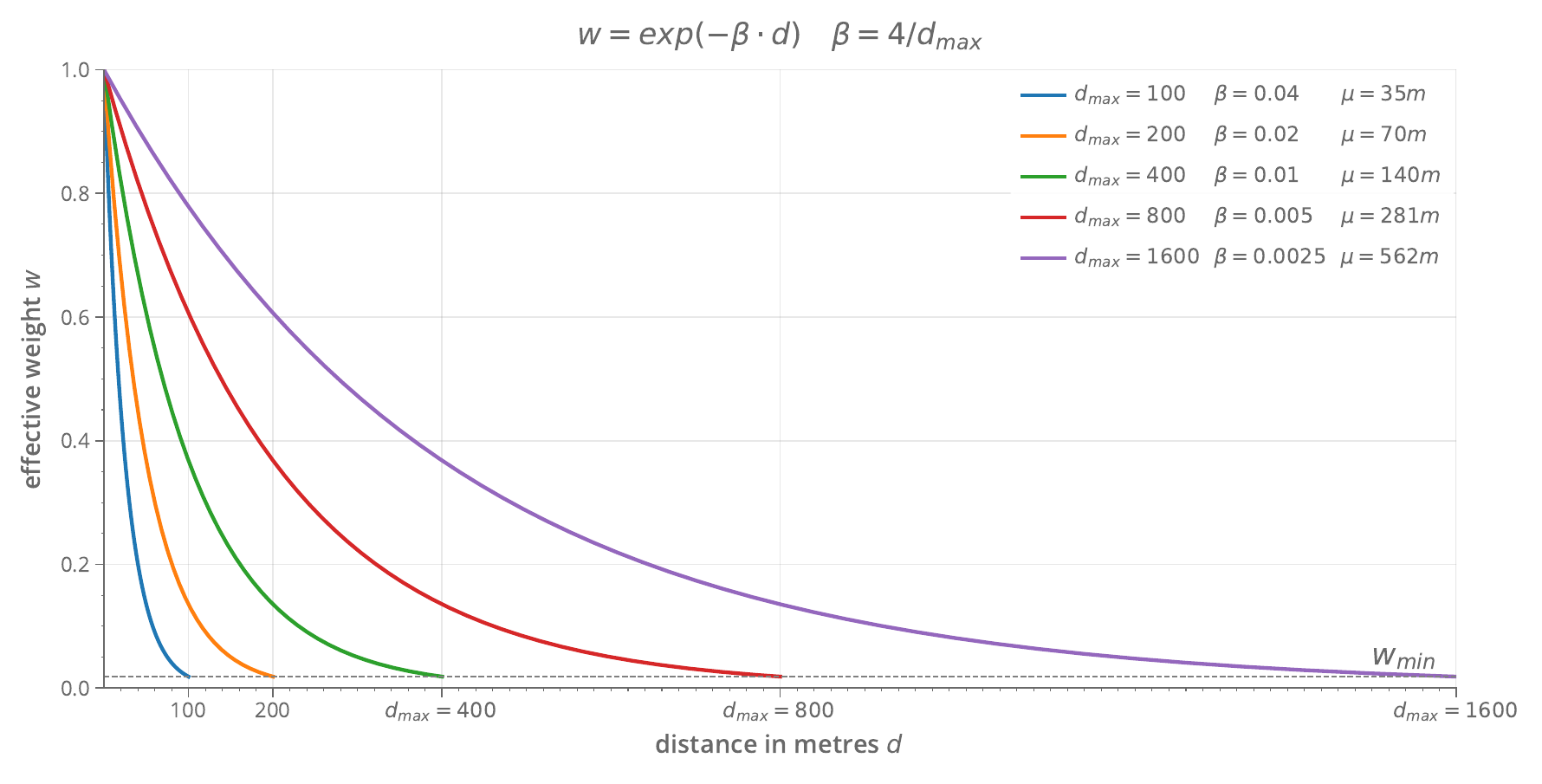}
 \caption[Spatial impedance curves for different $\beta$ parameters.]{Spatial impedance curves for different $\beta$ parameters. Nearer locations can be weighted more heavily than farther locations through use of the negative exponential decay function.}\label{fig:beta_decays_3}
\end{figure}

Gravity measures are inherently localised and, unlike windowed network measures, do not strictly require distance cutoffs. It is nevertheless computationally advantageous to retain thresholds commensurate with distances at which the decay renders additional computation sufficiently negligible; for the proceeding discussion, a range of impedances (Figure~\ref{fig:beta_decays_3}) is applied with the respective $\beta$ parameters anchored to distance thresholds $d_{max}$ through $\beta = 4 / d_{max}$. For example, a $100m$ distance threshold $d_{max}$ corresponds to $\beta=0.04$ and gives an average trip distance of $35m$. The distances and the corresponding impedances are selected with pedestrians in mind, where, for example, $\beta=0.005$ (800m $d_{\max}$) may represent relatively typical walking distances to bus stops \parencite{Sevtsuk2016, Handy1997}. Realistically, these values vary significantly based on the purpose of the trip because pedestrians may be willing to walk much farther for purposes such as recreation and fitness than for purposes such as shopping \parencite{Iacono2008}. These parameters also vary based on location: it can be argued that North American contexts are more walking suppressive \parencite{Reyer2014} than European equivalents. Regardless, pedestrians tend to be unwilling to walk distances greater than a mile ($1600m$, $\beta\approx0.0025$) for non-recreational purposes and, per the exponential decay function, are more likely to walk to nearer locations than those farther away \parencite{Baradaran2001, Scheurer2007, Duncan2011, Handy2001, Harris2001, Bates2007}.

As with closeness centrality, it may be preferable to use the gravity index in a continuous form when addressing varied intensities of nodes, poor quality network representations, or non-simplified networks. In this case, gravity is computed as the area under the curve for all reachable segments $S$ for the respective lower and upper segment bounds $a$ and $b$ at the specified impedance $\beta$:
\begin{equation}\label{eq:gravity_continuous}
 continuous\ G_{(i)} = \sum_{(a, b)}^{S} \int_{a}^{b}\ f(x)\ dx = \sum_{(a, b)}^{S}\ \frac{\exp(-\beta\cdot b) -\exp(-\beta\cdot a)}{-\beta}\, .
\end{equation}

\subsection{Betweenness centrality}

The shortest path from any node $j$ to any other node $k$ will pass through an assortment of nodes $i$, that is, unless $j$ and $k$ are directly adjacent. Betweenness 
\begin{equation}\label{eq:betweenness}
 Betweenness_{(i)} = \sum_{j\neq{i}} \sum_{k\neq{j}\neq{i}} n(j, k)\ i
\end{equation}
is the summation of shortest paths between all $(j, k)$ pairs of nodes passing through a given node $i$ \parencite{Freeman1977}, and in the case of street networks conveys how likely a street is to be traversed by people travelling between other locations.

Topological networks, which count the number of topological steps instead of summing distance weights, will frequently have multiple equally shortest paths for a particular $(j, k)$ pair. In these cases, the proportion of routes passing through node $i$ is used instead, such that if there were two shortest paths, then the sum is incremented by $\frac{1}{2}$ for the two respective routes, and likewise by $\frac{1}{3}$ if there were three. In the case of street networks weighted by distance or angular change, it is unlikely for two exactly equidistant shortest paths to exist, and this step can often be disregarded. An exception would be cities with heavily gridded layouts such as Manhattan or Barcelona, in which case routes within a tolerance of similarity could be considered equidistant, and therefore with proportional weightings used instead.

Localised betweenness, similar to localised Closeness, only considers other $(j, k)$ node pairs within the threshold cutoff distances. Unlike Closeness, global formulations applied to windowed graph analysis do not introduce unexpected behaviour.

For global betweenness measures, normalisation is beneficial unless there is a consistent methodological process for how the network boundaries were derived from the topological structure of the network, such as the network percolation based methods described in \textcite{Arcaute2016}. The boundaries would otherwise be arbitrary, and the number of shortest paths passing through a node would depend significantly on the extents of the chosen boundaries. However, for windowed betweenness, the distance thresholds are, in effect, already a form of implicit normalisation, and further use of explicit normalisation based on the parent-graph would prevent comparisons across locations or distance thresholds. Similar to windowed closeness measures, it would also lead to counter-intuitive results due to differing quantities of nodes.

Whereas Closeness and the gravity accessibility index directly take distances into account, betweenness only counts the number of shortest paths passing through a location. Betweenness can, however, be weighted by distances:
\begin{equation}\label{eq:weighted_betweenness}
 Betweenness_{(i)}=\sum_{j\neq{i}}\sum_{k\neq{j}\neq{i}} n(j, k)\ i\cdot\exp(-\beta\cdot d)\, ,
\end{equation}
in which case the negative exponential (see~\ref{eq:gravity}) reflects the notion that trips between closely located $(j, k)$ node-pairs are more likely to occur than those located farther apart, with $d$ in this case representing the corresponding trip distance for a given $(j, k)$ pair of nodes passing through node $i$.

\subsection{Network cycles}\label{network-cycles}

The previously mentioned measures are useful for exploring the role of shortest and simplest (angular) paths in network structures, but the richness of potential route choice options remains implicit. It is therefore interesting to consider the use of network cycles, simply a count of reachable cycles 
\begin{equation}
 Network\ Cycles_{(i)} = \sum_{j\neq{i}}^{cycles} 1
 \label{eq:network_cycles}
\end{equation}
to explicitly explore the potential complexity of available route options in a sense articulated by \textcite{Jacobs1961}, who argued that street configurations should permit `pools of cross-use' by providing a variety of overlapping route options that can be `mixed and mingled' within pedestrian walking thresholds. Cycles are analogous to a route encircling a block: setting off in one direction permits circling back towards the same location or mixing and matching many potential route configurations at intersections. Cycles reflect the potential complexity of interleaving route options as may be typified by gridded layouts in contrast to suburban cul-de-sacs designed to minimise such cycles. One drawback to the network cycles measure is that it does not work well for smaller distance thresholds at which cycles are precluded.

\section{Exploring centrality measures with the cityseer-api package}

The following analysis is an exploratory comparison of local network centralities, which are correlated to local mixed-uses and local land-use accessibilities with the intention of understanding which centralities relate most intuitively to notions of pedestrian-accessible land-uses. The network dataset consists of the full street network for Greater London as derived from the UK \emph{Ordnance Survey} \emph{Open Roads} dataset in both the primal and dual representations. The primal network consists of 192,843 nodes, with the street network buffered to the maximum distance threshold of $8km$ to avoid perturbations due to edge-effects \parencite{Gil2017}. Land-uses are derived from the UK \emph{Ordnance Survey} \emph{Points of Interest} (POI) dataset consisting of 388,323 points with Eastings and Northings for each location. These are classed according to a schema consisting of 617 categories and are assigned to the network nodes on either side of the closest adjacent street edge. Land-use accessibility and mixed-use measures are computed with the \code{cityseer-api} package, affording the use of dynamic land-use aggregation workflows based on distances computed directly over the street network while taking the direction of approach into account. These workflows avoid larger-scaled gridded or zonal aggregations of land-uses, which otherwise incur a loss of resolution while inducing a disassociation between the observed centralities and land-uses for a given street-front location. These methods are described at length in
\ifpaper \textcite{Simons2021d} and \textcite{Simons2021b}
\else Section~\ref{localised-land-use-diversity-methods} and Section~\ref{the-cityseer-api-Python-package}.
\fi

The correlation plots make use of Spearman Rank $\rho$ correlation coefficients (step-wise monotonicity of the data) as opposed to Pearson's $r$ (linearity of the data) because heavily skewed datasets, as is the case with betweenness measures, would otherwise require aggressive preprocessing steps (e.g.~max-log optimised boxcox transformations). The observations must be framed with the caveat that no network measure should be expected to correlate perfectly to land-uses because there are broader considerations at work, including population densities, the nature of the building fabric, affordability, land-use regulations, etc. It should also be noted that correlations tend to be boosted at larger distance thresholds because increasing aggregation levels tend to smooth the variance of the data. This phenomenon is a characteristic of the Modifiable Areal Unit Problem \parencite{Fotheringham1991} with the implication that correlations are not directly equivalent across different scales of aggregation.

\begin{figure}[htbp]
 \centering
 \includegraphics[width=\textwidth, keepaspectratio]{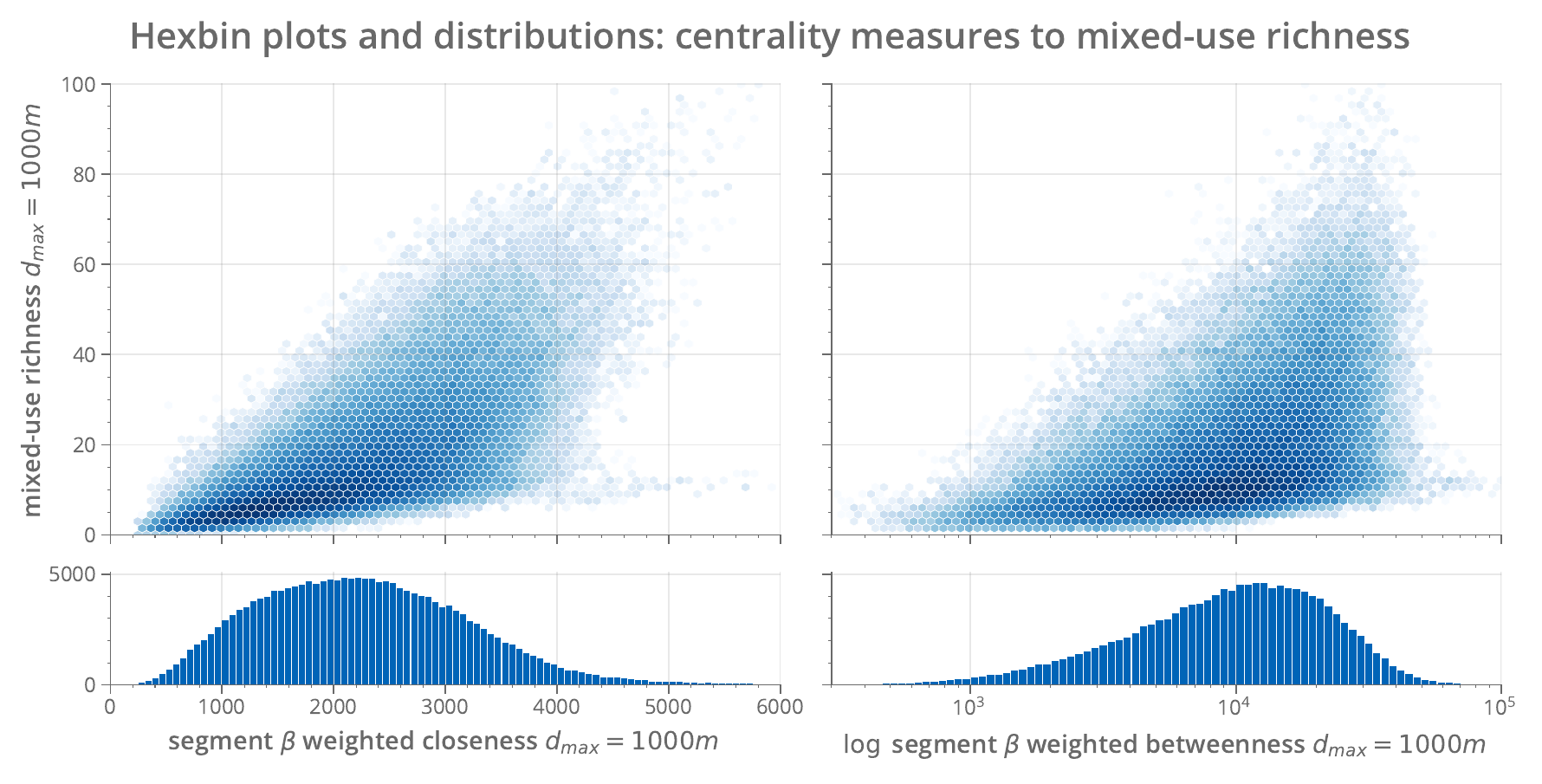}
 \caption[Distributions and hexbin plots for gravity and weighted betweenness against mixed-uses.]{Example distributions and hexbin plots for gravity and weighted betweenness against mixed-uses. Closeness related measures (such as gravity and node densities) show heavier distributions at smaller values with higher values reflecting a relatively linear increase in relation to mixed-uses. Betweenness measures assume a more ambiguous relationship.}\label{fig:primal_hexbin_gravity_betweenness}
\end{figure}

\begin{figure}[htbp]
 \centering
 \includegraphics[width=0.55\textwidth, keepaspectratio]{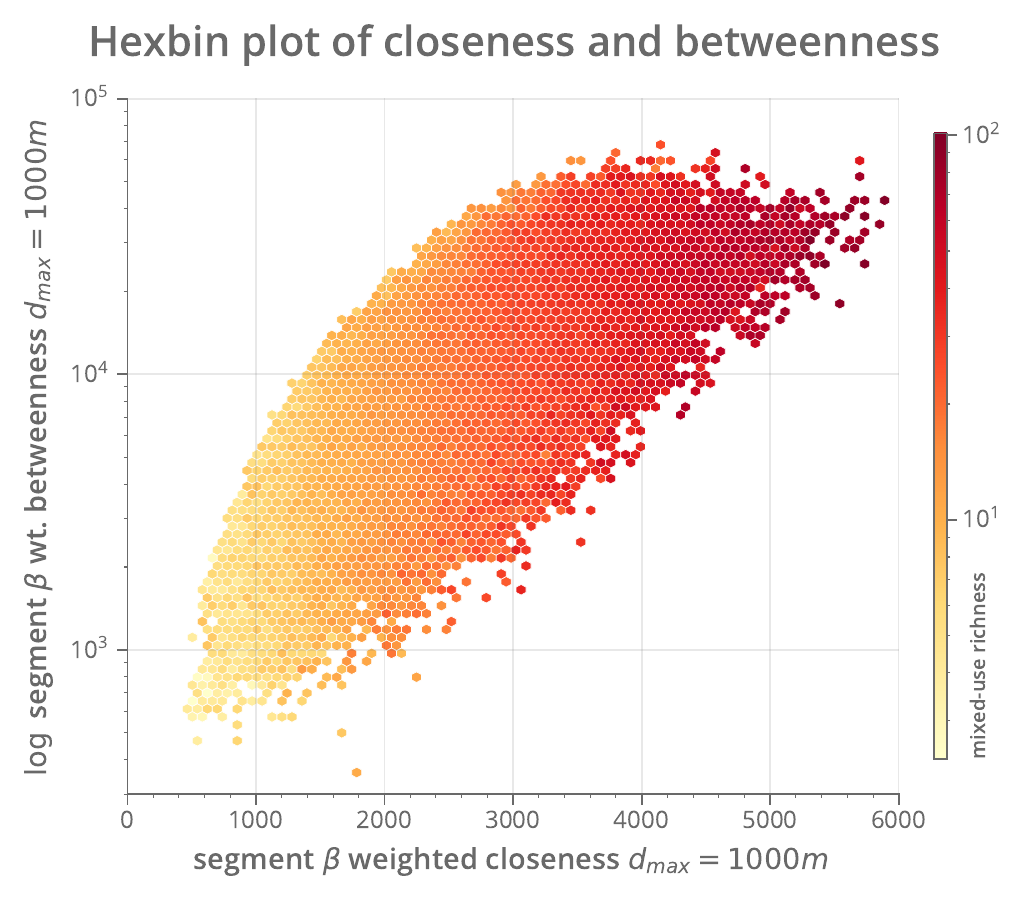}
 \caption[Hexbin plot of gravity, weighted betweenness, and mixed-uses.]{Hexbin plot showing gravity and weighted betweenness, with higher degrees of mixed-uses represented by increasing intensities of colour. Higher mixed-uses associate with a higher degree of closeness, with the relationship less well defined for betweenness.}\label{fig:primal_gravity_betweenness_mixed_hexbin}
\end{figure}

\begin{figure}[p]
 \centering
 \includegraphics[width=\textwidth, height=0.975\textheight, keepaspectratio]{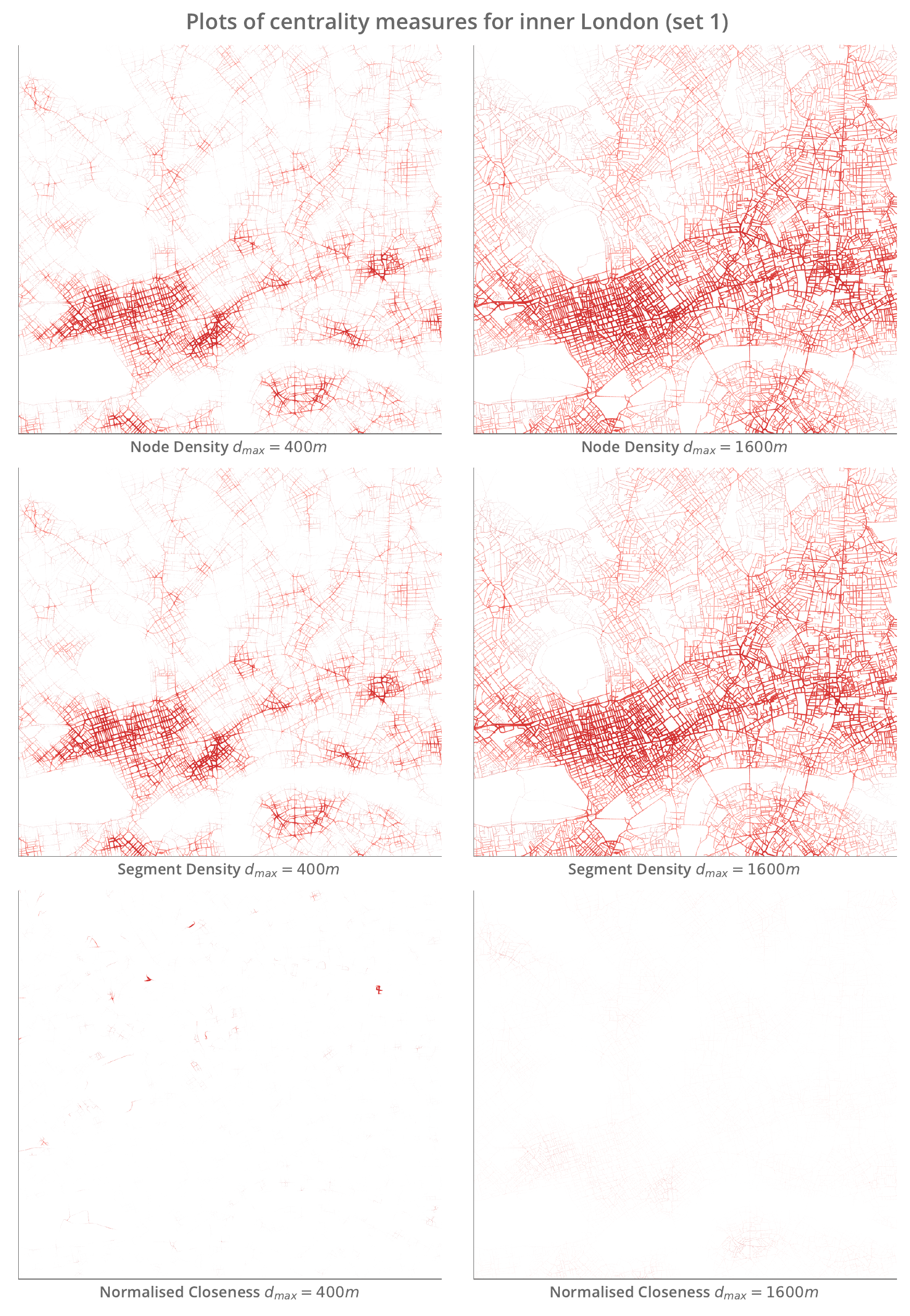}
 \caption[Comparative plots of primal centrality measures for inner London (part 1).]{Comparative plots of selected primal centrality measures for inner London (part 1). Distance thresholds as indicated. Normalised Closeness is included for the purpose of demonstrating the difficulty with using this measure for localised forms of network analysis.}\label{fig:primal_centrality_comparisons_contrast}
\end{figure}

\begin{figure}[p]
 \centering
 \includegraphics[width=\textwidth, height=0.975\textheight, keepaspectratio]{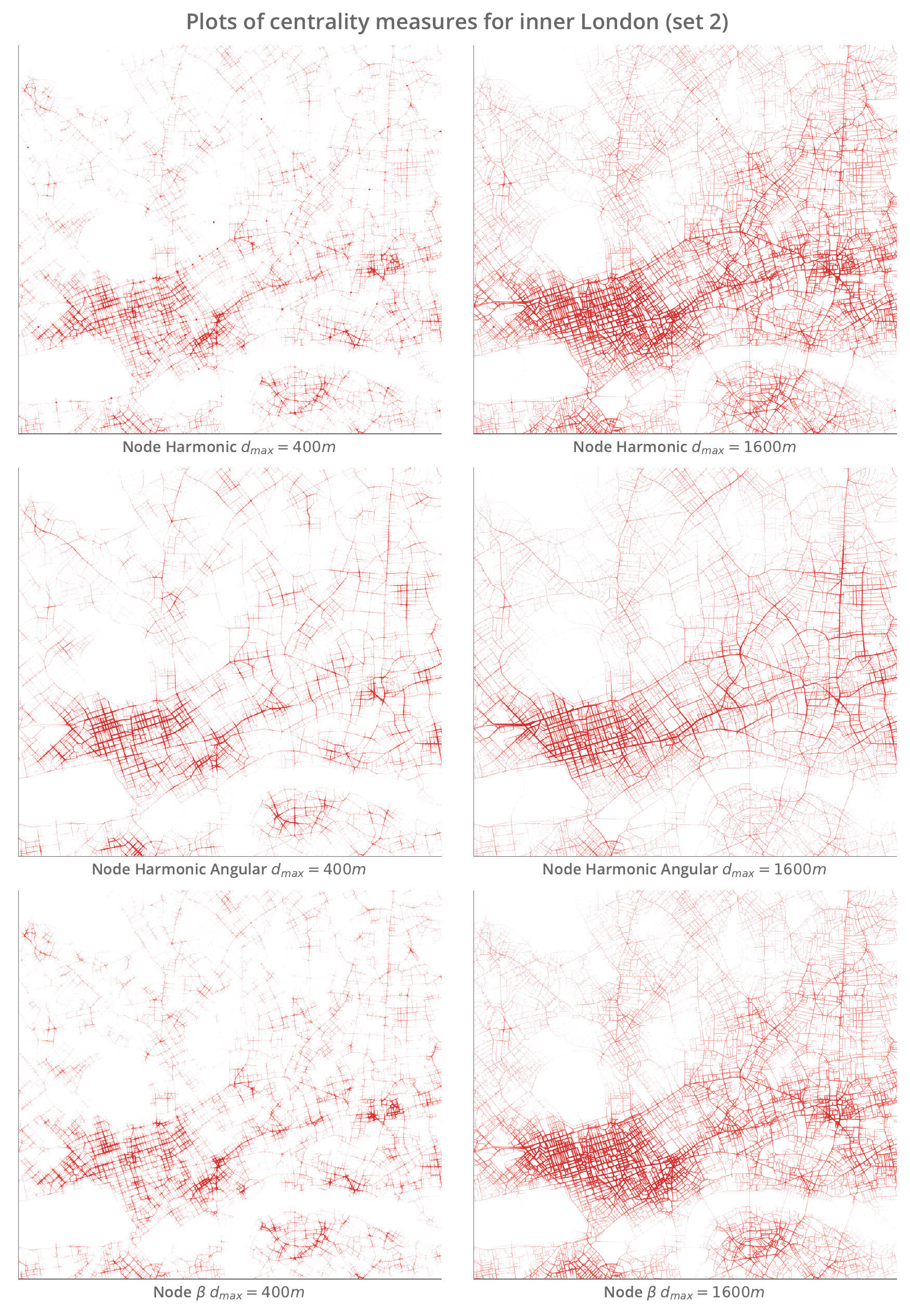}
 \caption[Comparative plots of primal centrality measures for inner London (part 2).]{Comparative plots of selected primal centrality measures for inner London (part 2). Distance thresholds as indicated.}\label{fig:primal_centrality_comparisons_closeness_node}
\end{figure}

\begin{figure}[p]
 \centering
 \includegraphics[width=\textwidth, height=0.975\textheight, keepaspectratio]{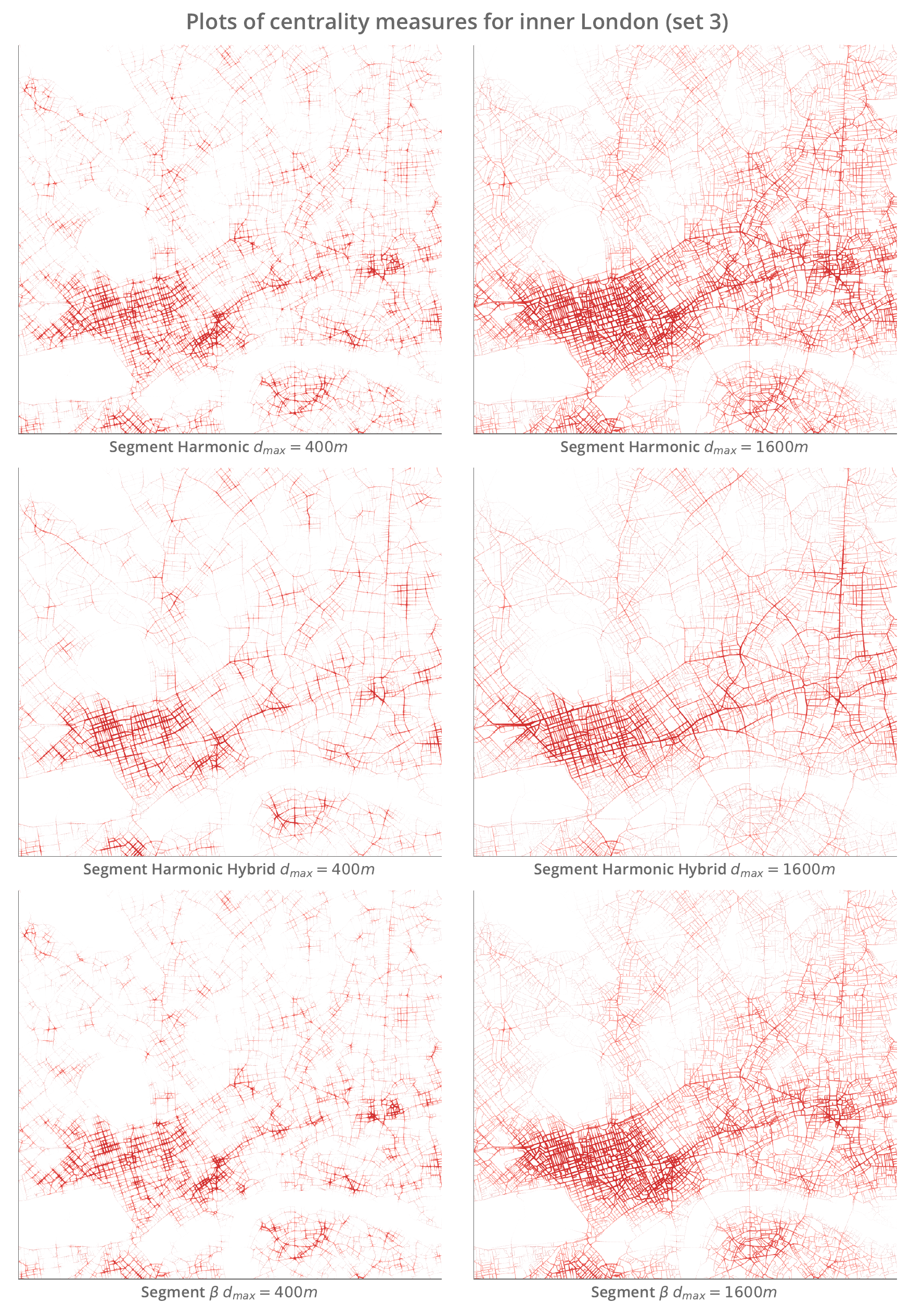}
 \caption[Comparative plots of primal centrality measures for inner London (part 3).]{Comparative plots of selected primal centrality measures for inner London (part 3). Distance thresholds as indicated.}\label{fig:primal_centrality_comparisons_closeness_segment}
\end{figure}

\begin{figure}[p]
 \centering
 \includegraphics[width=\textwidth, height=0.975\textheight, keepaspectratio]{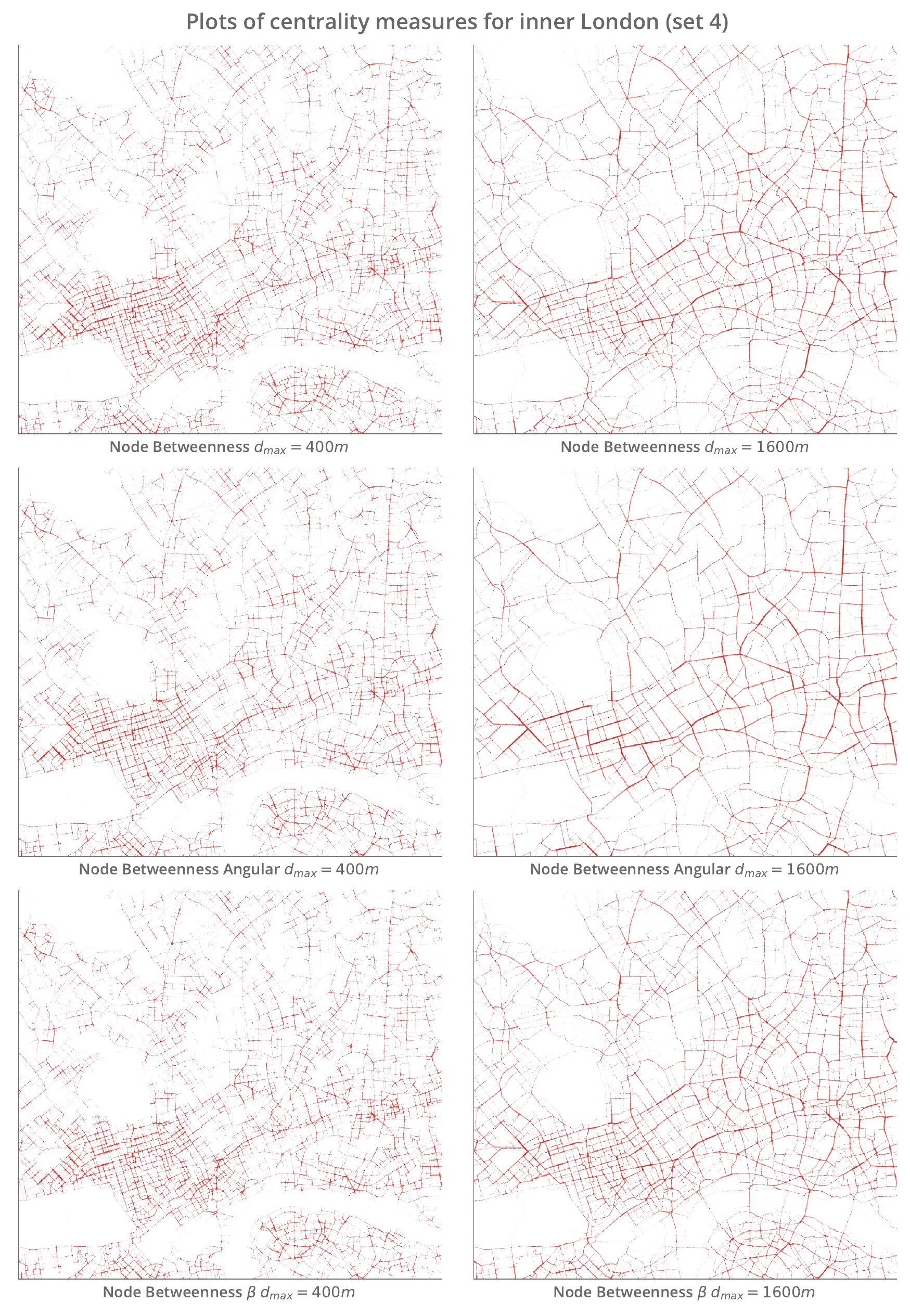}
 \caption[Comparative plots of primal centrality measures for inner London (part 4).]{Comparative plots of selected primal centrality measures for inner London (part 4). Distance thresholds as indicated.}\label{fig:primal_centrality_comparisons_betweenness_node}
\end{figure}

\begin{figure}[p]
 \centering
 \includegraphics[width=\textwidth, height=0.975\textheight, keepaspectratio]{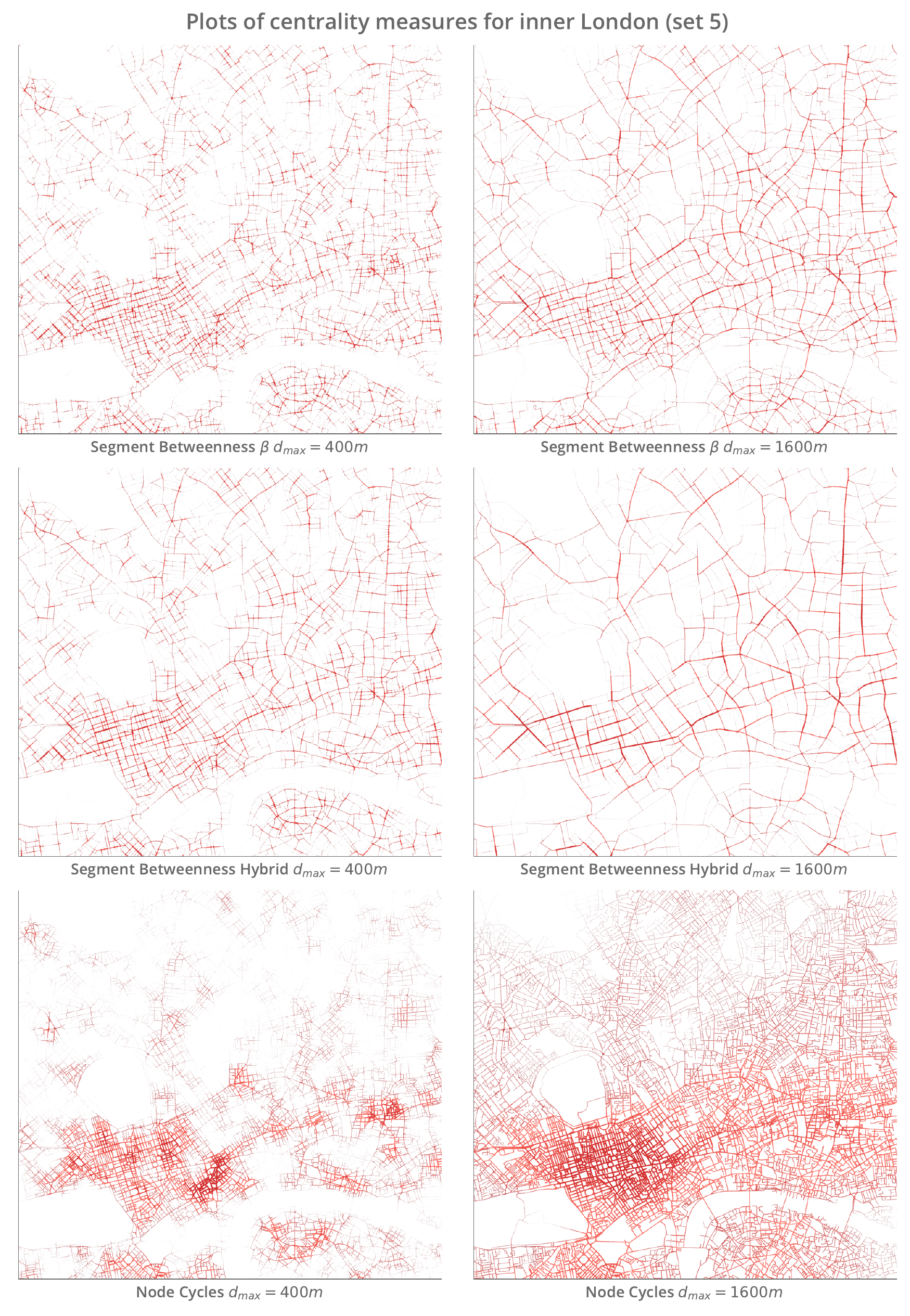}
 \caption[Comparative plots of primal centrality measures for inner London (part 5).]{Comparative plots of selected primal centrality measures for inner London (part 5). Distance thresholds as indicated.}\label{fig:primal_centrality_comparisons_betweenness_segment}
\end{figure}

A range of centralities is computed from $50m$ to $8km$ with a visual survey of the $400m$ and $1600m$ cases plotted for a selection of measures in the full page plots from Figures~\ref{fig:primal_centrality_comparisons_contrast} through~\ref{fig:primal_centrality_comparisons_betweenness_segment}.

Figure~\ref{fig:primal_hexbin_gravity_betweenness} shows that closeness-based measures assume a heavier distribution at smaller values with higher values assuming a relatively linear increase in relation to the presence of mixed-uses. Betweenness is more uncertain, with upper outliers associated with both lower and higher values of mixed-uses, perhaps reflecting the intuition that large volumes of through-traffic may in some cases be accompanied by a high quantity and variety of land-uses, such as high-streets, whereas being devoid thereof in other cases, such as motorways. This observation is reinforced in Figure~\ref{fig:primal_gravity_betweenness_mixed_hexbin}, where the most intensive levels of mixed-uses tend to increase with closeness while being more loosely associated with betweenness.

\begin{figure}[htbp]
 \centering
 \includegraphics[width=\textwidth, height=0.425\textheight, keepaspectratio]{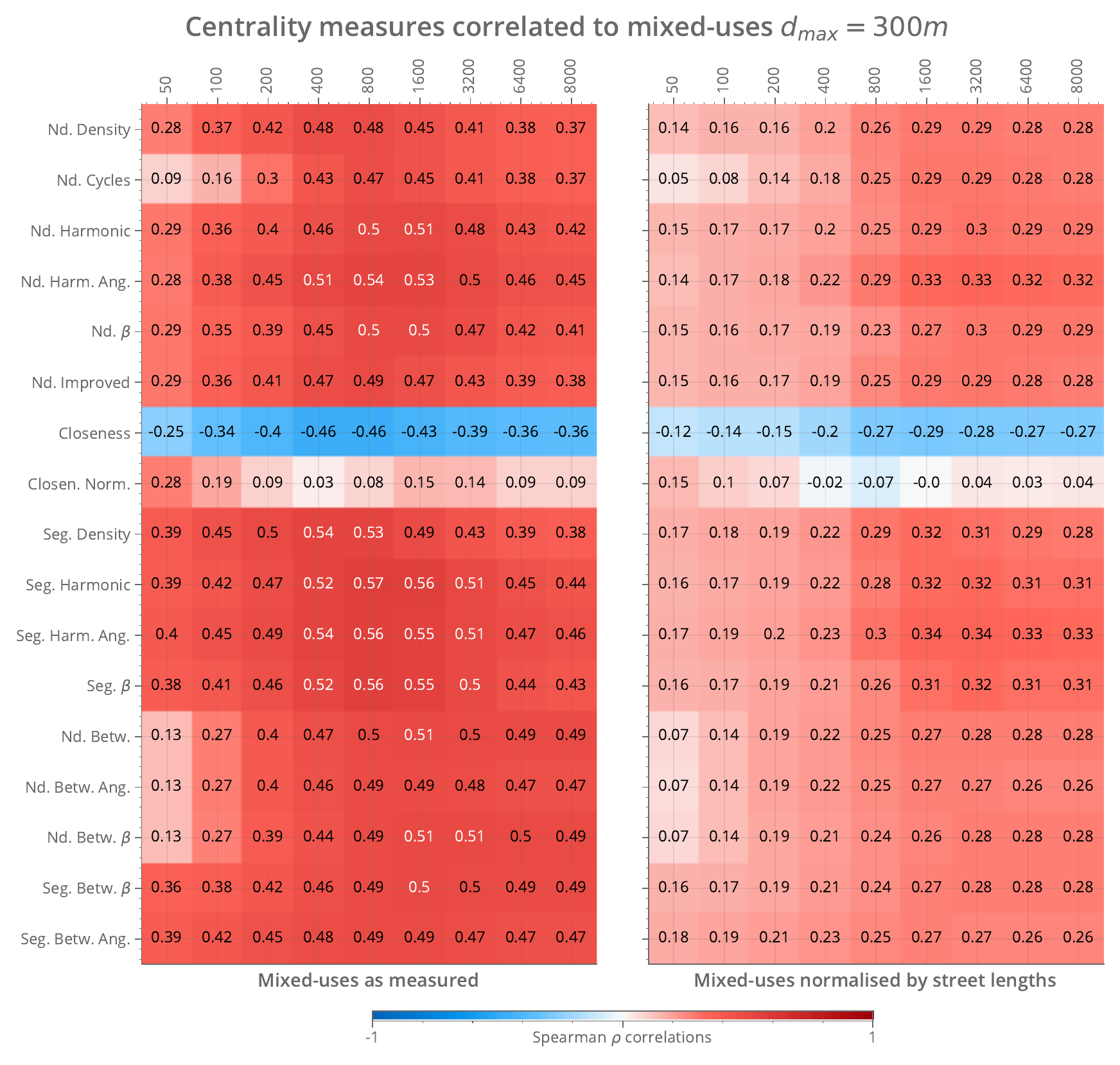}
 \caption[Correlation grids comparing network centralities to local mixed-uses.]{Correlation grids comparing network centralities to distance weighted mixed-uses $\beta=0.01\bar{3}$ / $d_{\max}=300m$ $Hill_{q=0}$ as a proxy for high streets. Length normalised correlations normalise the number of mixed-uses by street lengths.}\label{fig:primal_centralities_corr_grid_mu_300}
\end{figure}

\begin{figure}[htbp]
 \centering
 \includegraphics[width=\textwidth, height=0.425\textheight, keepaspectratio]{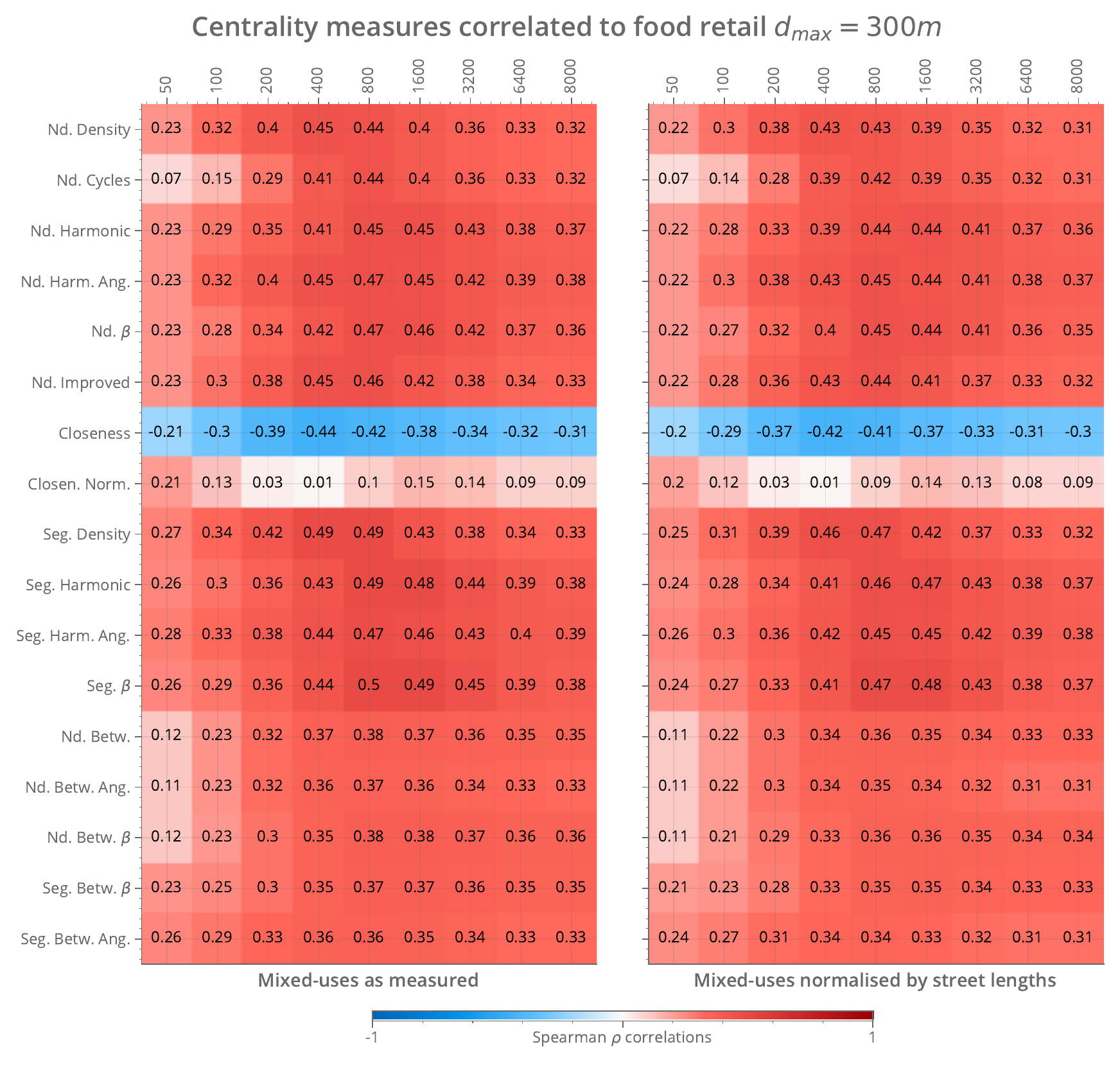}
 \caption[Correlation grids comparing network centralities to local food retail accessibility.]{Correlation grids comparing network centralities to weighted local food retail accessibility $d_{\max}=300m$ as a proxy for high streets. Length normalised correlations normalise the number of land-uses by street lengths.}\label{fig:primal_centralities_corr_grid_retail_300}
\end{figure}

\begin{figure}[htbp]
 \centering
 \includegraphics[width=\textwidth, height=0.425\textheight, keepaspectratio]{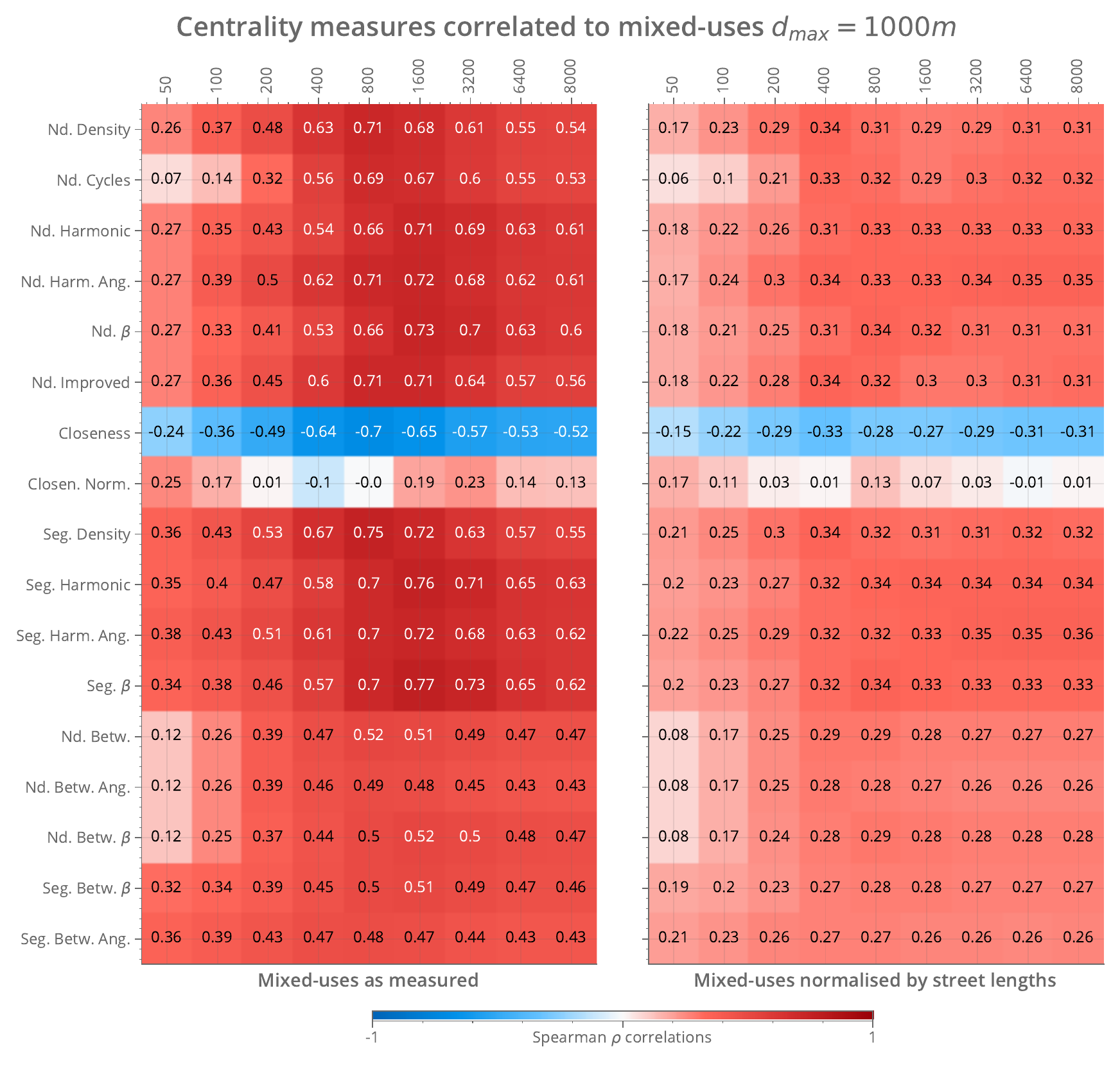}
 \caption[Correlation grids comparing network centralities to neighbourhood scale mixed-uses.]{Correlation grids comparing network centralities to mixed-uses $\beta=0.004$ / $d_{\max}=1000m$ $Hill_{q=0}$ as a proxy for vibrant districts. Length normalised correlations normalise the number of mixed-uses by street lengths.}\label{fig:primal_centralities_corr_grid_mu_1000}
\end{figure}

\begin{figure}[htbp]
 \centering
 \includegraphics[width=\textwidth, height=0.425\textheight, keepaspectratio]{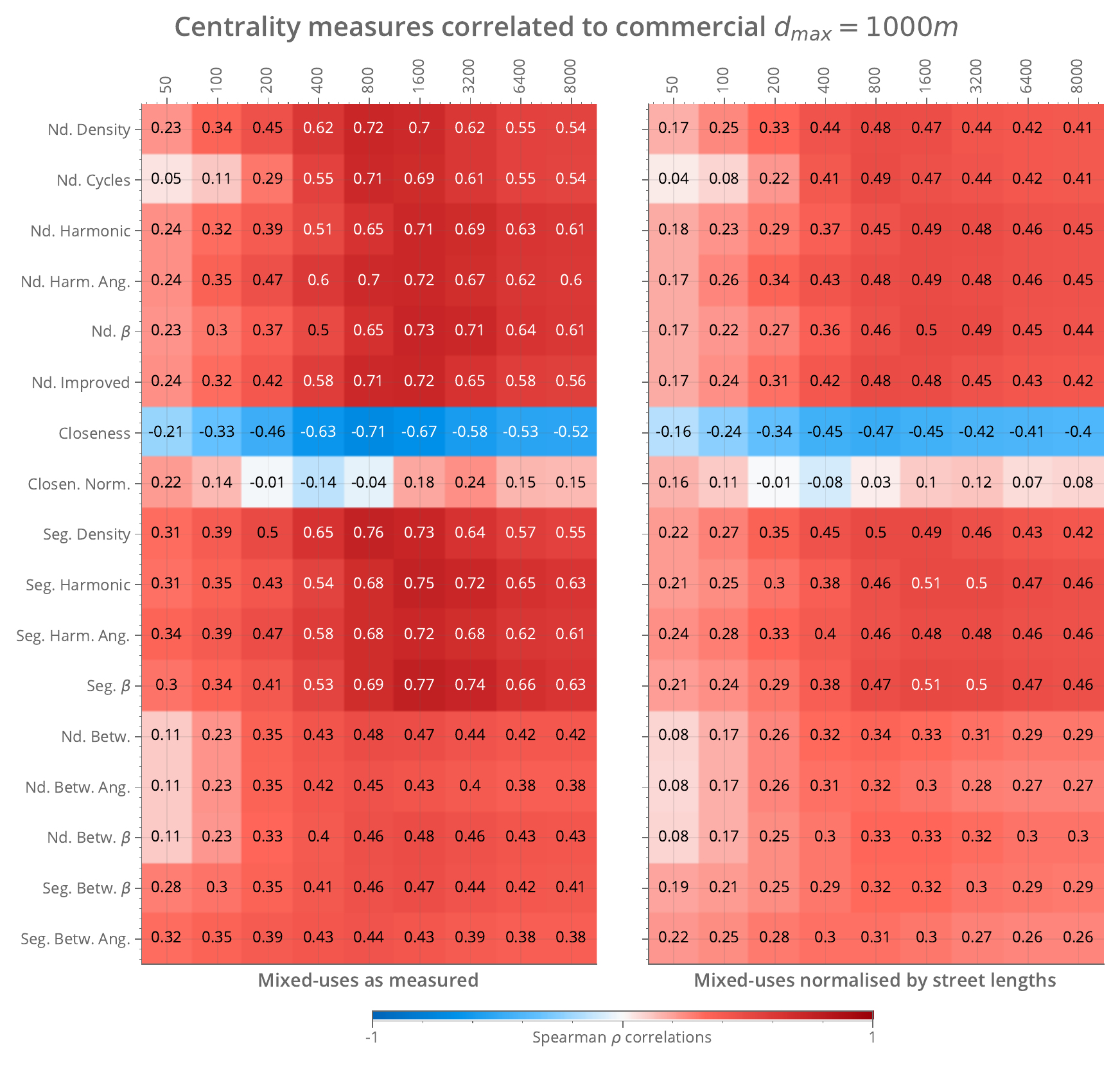}
 \caption[Correlation grids comparing network centralities to neighbourhood scale commercial accessibility.]{Correlation grids comparing network centralities to commercial land-use accessibility $d_{\max}=1000m$ as a proxy for vibrant districts. Length normalised correlations normalise the number of land-uses by street lengths.}\label{fig:primal_centralities_corr_grid_commercial_1000}
\end{figure}

\begin{figure}[htbp]
 \centering
 \includegraphics[width=\textwidth, height=0.425\textheight, keepaspectratio]{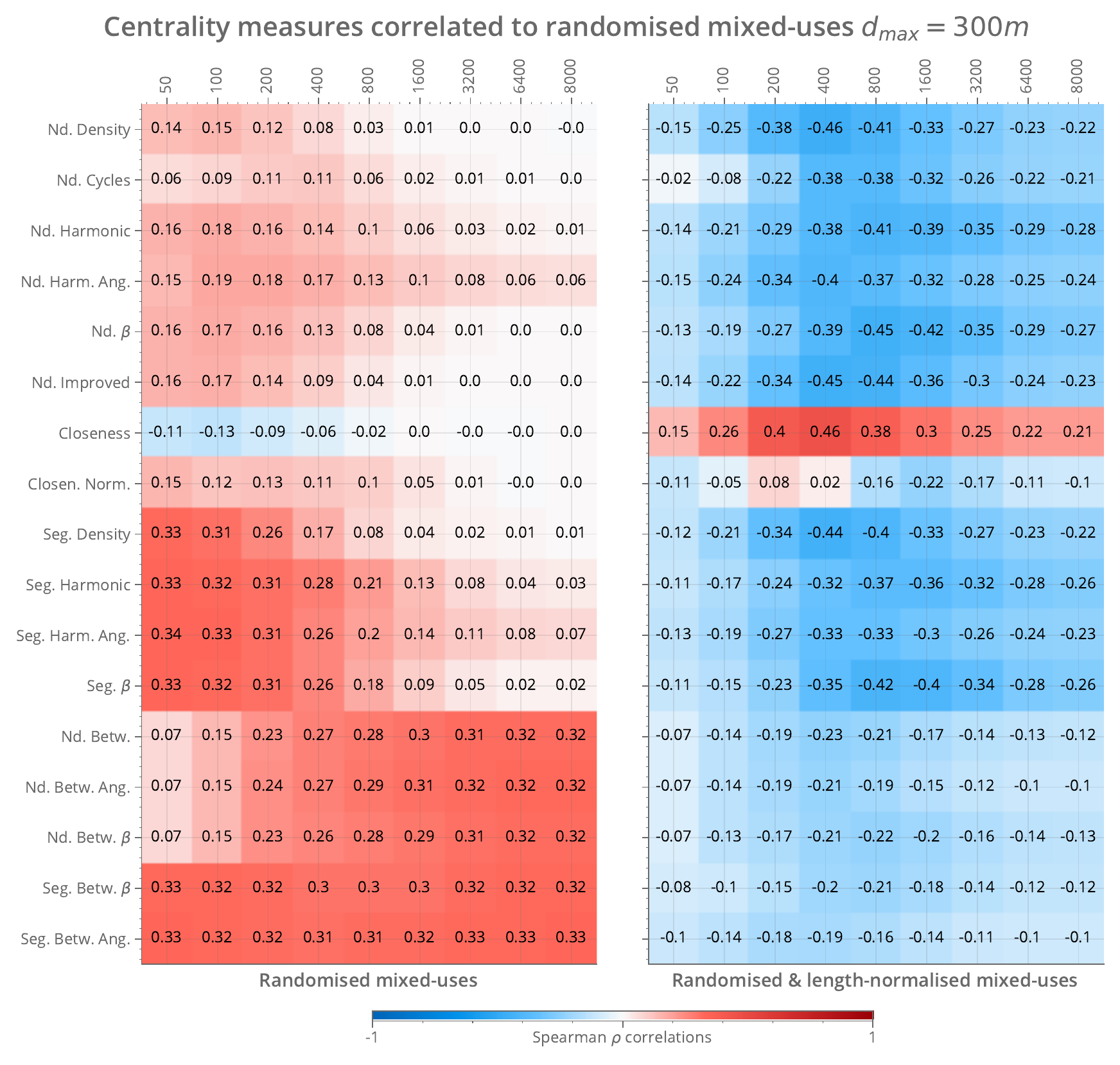}
 \caption[Correlation grids comparing network centralities to \textbf{randomised} local mixed-uses.]{Correlation grids comparing network centralities to \textbf{randomised} local mixed-uses $\beta=0.01\bar{3}$ / $d_{\max}=300m$ $Hill_{q=0}$. This plot is included for the purpose of demonstrating how access to mixed-uses can increase purely as a function of the network (left) and how this can be controlled for by normalising by street lengths (right).}\label{fig:primal_centralities_corr_grid_mu_300_rdm}
\end{figure}

\begin{figure}[htbp]
 \centering
 \includegraphics[width=\textwidth, height=0.425\textheight, keepaspectratio]{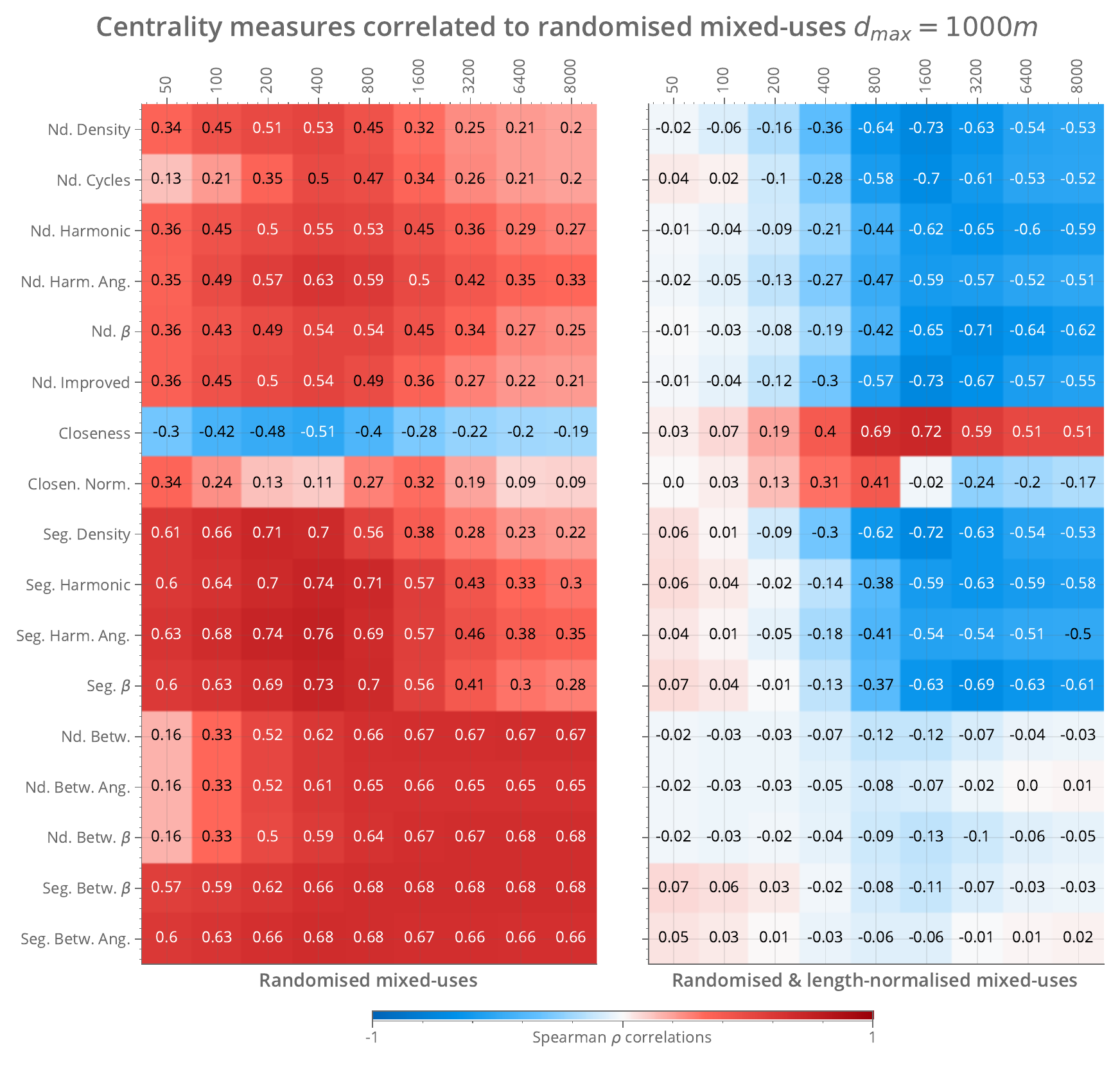}
 \caption[Correlation grids comparing network centralities to \textbf{randomised} neighbourhood scale mixed-uses.]{Correlation grids comparing network centralities to \textbf{randomised} neighbourhood scale mixed-uses $\beta=0.004$ / $d_{\max}=1000m$ $Hill_{q=0}$. This plot is included for the purpose of demonstrating how access to mixed-uses can increase purely as a function of the network (left) and how this can be controlled for by normalising by street lengths (right).}\label{fig:primal_centralities_corr_grid_mu_1000_rdm}
\end{figure}

The subsequent correlation plots are computed against distance weighted mixed-use richness ($Hill_{q=0}$) and Food Retail accessibility at the smaller distance threshold of $\beta=0.01\bar{3}$ / $d_{\max}=300m$ as a proxy for high streets (see Figure~\ref{fig:primal_centralities_corr_grid_mu_300} and Figure~\ref{fig:primal_centralities_corr_grid_retail_300}), with distance-weighted mixed-use richness ($Hill_{q=0}$) and Commercial accessibilities used at the larger distance threshold of $\beta=0.004$ / $d_{\max}=1000m$ as a proxy for wider mixed-use districts (see Figure~\ref{fig:primal_centralities_corr_grid_mu_1000} and Figure~\ref{fig:primal_centralities_corr_grid_commercial_1000}). Note that the use of weighting based on spatial impedances means that these measures are more locally focussed than the $d_{\max}$ values may otherwise imply. Each of these figures is shown as two correlation gridplots: the first representing correlations of centralities to raw mixed-use and land-use calculations while the second shows the same measures normalised by street lengths. The latter controls for network effects, which can be illustrated with the use of randomised land-use locations generated by randomly `scrambling' existing land-uses within the extents of Greater London; so doing, land-uses become more evenly distributed through space instead of the typically clustered land-use hot-spots observed for real-world scenarios. The measurement of randomised mixed-uses at the smaller distance threshold (Figure~\ref{fig:primal_centralities_corr_grid_mu_300_rdm}) would therefore invoke some peculiar results: streets ordinarily expected to have higher access to mixed-uses would instead have lower access and vice-versa, and the correlations are accordingly mostly negligible. The situation changes, however, once expanding the land-use aggregation thresholds to larger distances (Figure~\ref{fig:primal_centralities_corr_grid_mu_1000_rdm}) which now gain higher access to mixed-uses purely as a function of higher levels of network centrality. This effect of the network on the measured mixed-uses can be mitigated through street length normalisation. However, it should be noted that the non-normalised correlations remain a valid form of observation, albeit with the proviso that access to land-uses can increase both as a function of the intensity of land-uses per unit street-frontage or as a function of greater network centrality, which increases the amount of accessible street-frontage.

Generally: closeness-like measures (including density and $\beta$ variants) offer stronger correlations to mixed-uses for both the smaller high-street and larger neighbourhood-scale mixed-use scenarios; global variants of closeness are wholly ineffective and behave counter-intuitively; segmented versions offer stronger correlations than node-based versions, more so at smaller distances; and weighted measures offer similar or slightly better correlations than unweighted measures, though at equivalently larger distance thresholds. The $\beta$ weighted measures are distinguished by an explicit spatial impedance that weights nearer locations substantially more heavily than distant locations (Figure~\ref{fig:beta_decays_3}). The aggressive $\beta$ used at smaller distance thresholds causes a `lag' for correlations when compared to non $\beta$ measures at equivalent $d_{\max}$, but may ultimately achieve similar or higher correlations at larger thresholds while retaining a greater degree of local specificity than the distance thresholds might otherwise imply. Simplest-path angular methods confer an advantage for smaller distance centralities in the context of mixed-uses, though this advantage tends to dissipate when compared to retail and commercial accessibilities, particularly at the larger scale. This observation may correspond to the intuition that angular methods are suited to the accentuation of high-streets, which are often linear and located along thoroughfares through historic town or city centres. In contrast, more expansive mixed-use districts demonstrate a more agglomerative and fractal (space-filling) dynamic which is better suited to shortest paths. Interestingly, the comparatively crude node-density and segment-density measures perform similarly or better than the more sophisticated measures at the smallest distance thresholds. This observation changes at larger thresholds where the more refined measures take the advantage: it could be surmised that brute access to as many locations as possible remains an overriding consideration, particularly at smaller walking tolerances.

Note that two issues arise when comparing correlations at successively larger distance thresholds. The first is related to the Modifiable Areal Unit Problem~\parencite{Fotheringham1991}: the larger the threshold, the larger the aggregation and the more likely it is that smoother distributions with reduced variances will boost correlations. The second is that properties of smaller distance thresholds become subsumed by larger thresholds, with the implication that correlations already present at smaller scales may mask the weakening of correlations at the peripheries of larger thresholds. Against this backdrop, the majority of the measures achieve near to maximal correlations in the vicinity of $400m$--$1200m$ at which point the correlations tend to stagnate or begin to decrease as thresholds further increase, indicating that street segments sufficiently far away may start to muddy the relevancy of the centrality measure relative to the local land-use context.

\begin{figure}[htbp]
 \centering
 \includegraphics[width=\textwidth, height=0.4\textheight, keepaspectratio]{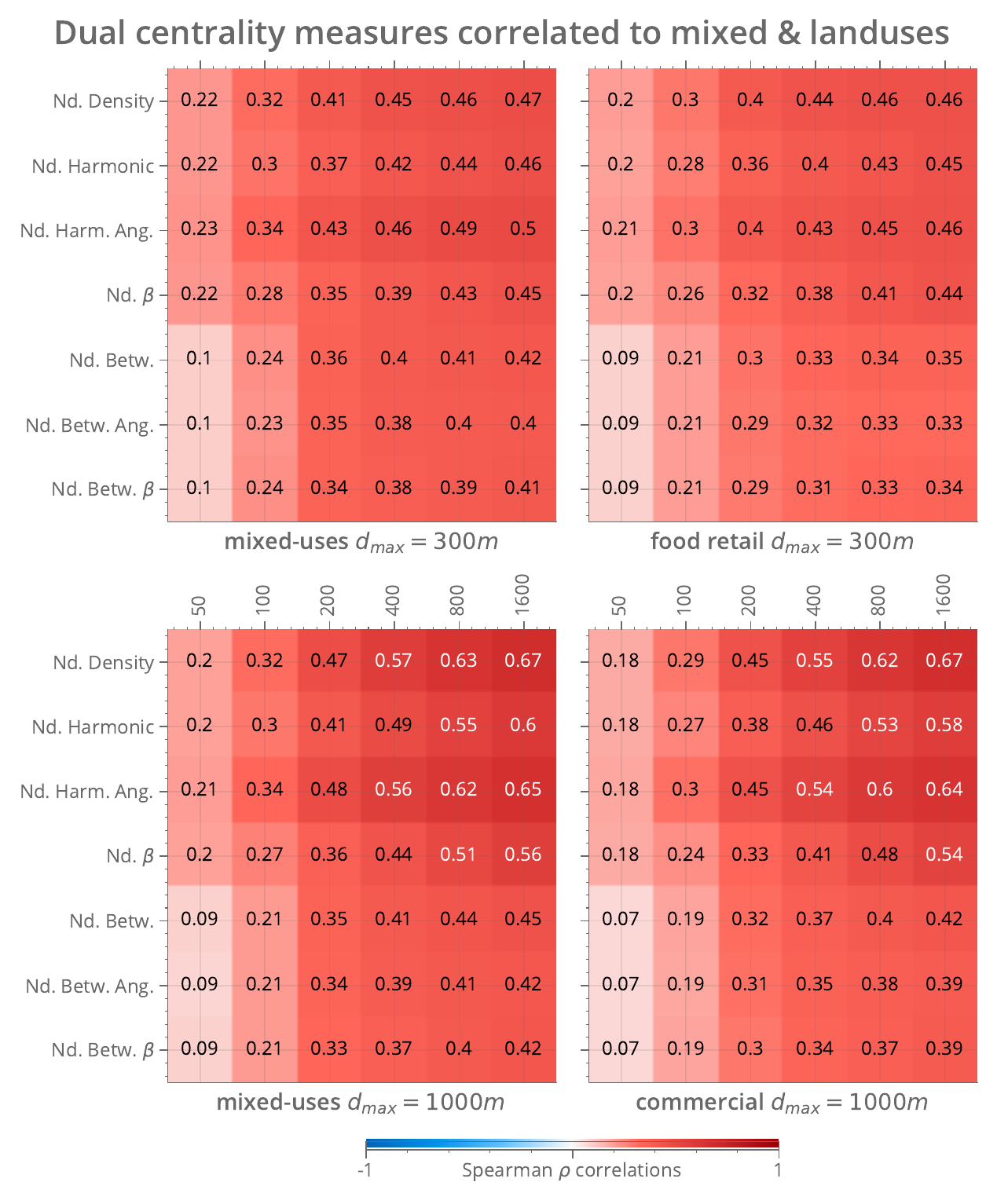}
 \caption[Correlation grids comparing dual network centralities to mixed-uses and land-use accessibilities.]{Correlation grids comparing dual network centralities to mixed-uses and land-use accessibilities on the dual network. Weighted mixed-use richness ($Hill_{q=0}$) and Food Retail at the smaller distance thresholds of $\beta=0.01\bar{3}$ / $d_{\max}=300m$ serve as a proxy for high streets, whereas weighted mixed-use richness ($Hill_{q=0}$) and Commercial accessibilities at the larger distance thresholds of $\beta=0.004$ / $d_{\max}=1000m$ serve as a proxy for wider mixed-use districts.}\label{fig:dual_centralities_corr_grid}
\end{figure}

\begin{figure}[htbp]
 \centering
 \includegraphics[width=\textwidth, height=0.6\textheight, keepaspectratio]{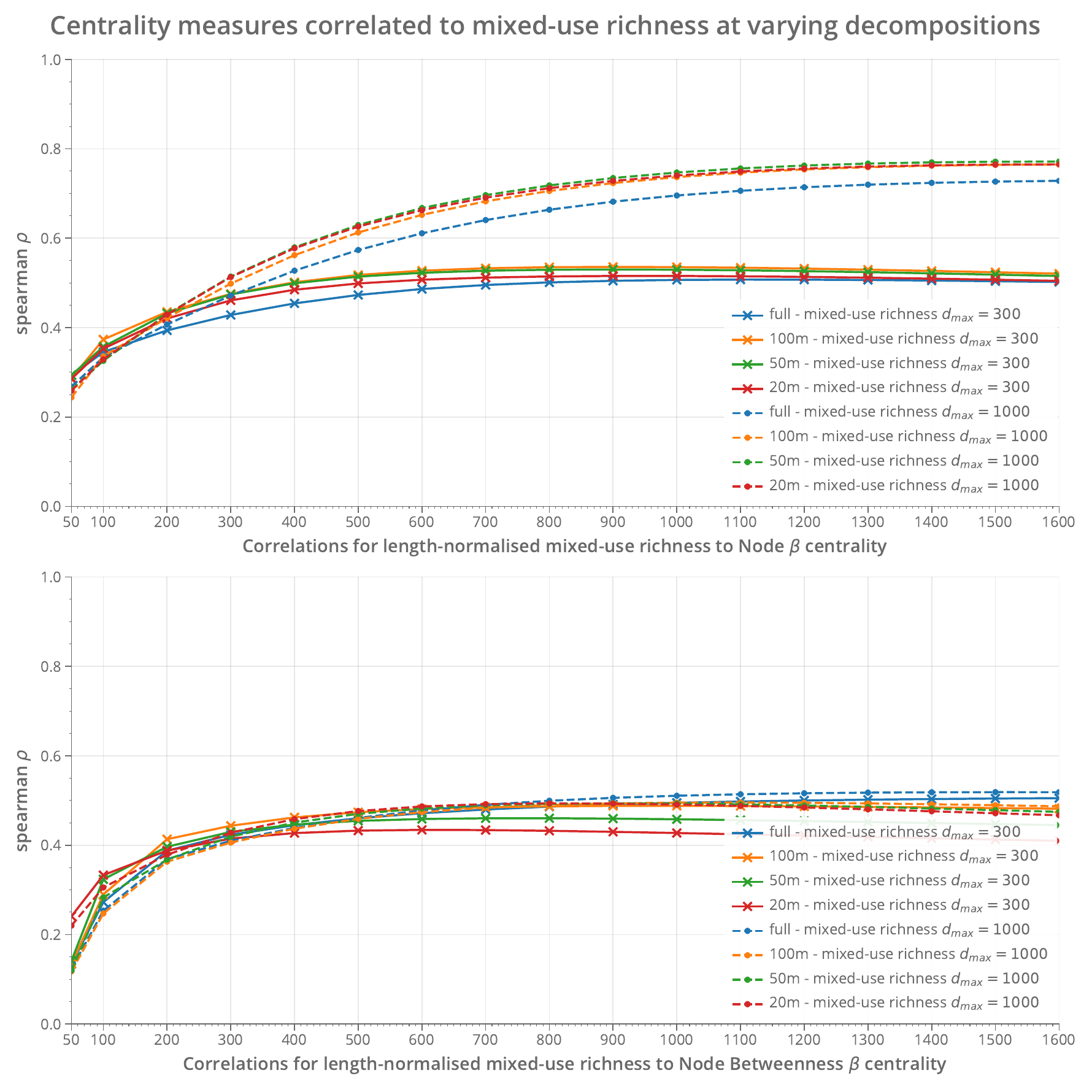}
 \caption[Correlations for shortest-path centrality measures to mixed-uses at varying levels of decomposition.]{Correlations for shortest-path network centrality measures to mixed-uses at varying levels of network decomposition. The respective network centrality measures are compared to mixed-use richness ($Hill_{q=0}$) at both smaller ($\beta=0.01\bar{3}$ / $d_{\max}=300m$) and greater ($\beta=0.004$ / $d_{\max}=1000m$) distance thresholds.}\label{fig:primal_decompositions}
\end{figure}

\begin{figure}[htbp]
 \centering
 \includegraphics[width=\textwidth, keepaspectratio]{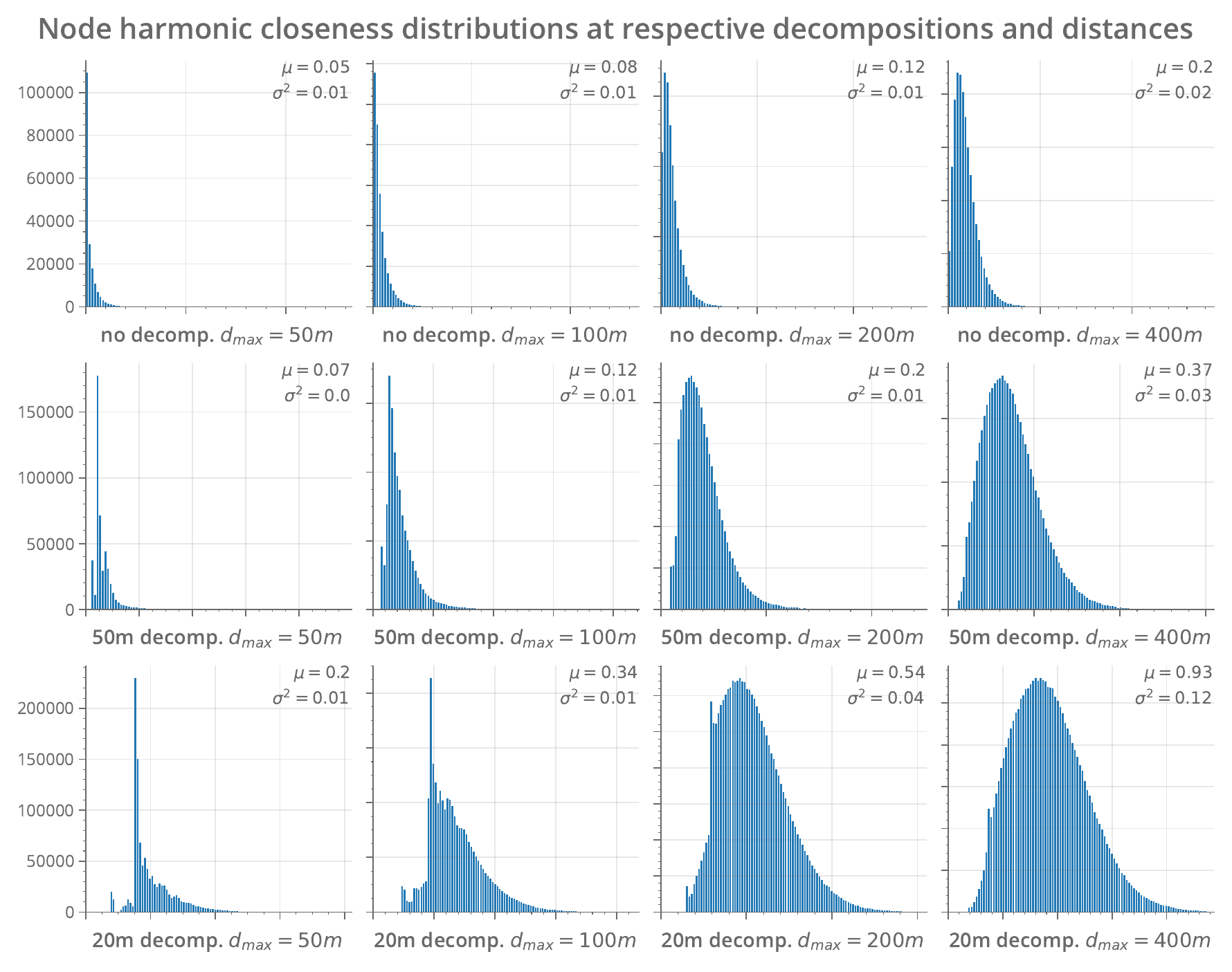}
 \caption[Comparative distributions at small distance thresholds (node based).]{Comparative distributions of node-based harmonic closeness at small distance thresholds thresholds, compared across different levels of decomposition.}\label{fig:distribution_degeneration_node_harm}
\end{figure}

Use of the dual network appears to infer no tangible advantage (see Figure~\ref{fig:dual_centralities_corr_grid}): correlations for the respective measures are similar though generally weaker for betweenness measures. A further drawback in using the dual topology is that segment-based methods cannot be used because this would incur repeated summations over duplicated segment lengths.

The preceding observations are based on the full topological network as derived from \emph{Ordnance Survey} \emph{Open Roads}; however, it is sometimes beneficial or necessary to sample centralities at more finely spaced intervals on the road network. The following considers the use of `decomposition' as a strategy
\ifpaper \parencite{Simons2021b}
\else (see Section~\ref{the-cityseer-api-Python-package})
\fi
whereby the road network is subdivided into smaller links (edges) based on a maximum distance. Decomposition is here applied at the $100m$, $50m$, and $20m$ levels on the primal network, with the latter case ultimately yielding 1,097,510 nodes. Decomposition of the network has a mixed impact on the strength of correlations, varying based on the degree of decomposition, the distance threshold, shortest or simplest path heuristic, and the chosen measures (here plotted for selection of measures in Figure~\ref{fig:primal_decompositions}). Gentler decomposition at $100m$ or $50m$ can boost correlations, whereas aggressive decomposition at $20m$ is less advantageous. Visual clues to this dynamic can be found in the distributions of the measures at small thresholds: correlations soften across-the-board for thresholds below $400m$ because the topology of any given network causes shortest-path algorithms to terminate increasingly abruptly as the search distance thresholds decrease, thus leading to artefacts in the distributions of the measures (see Figure~\ref{fig:distribution_degeneration_node_harm}). Accordingly, algorithms may terminate within one or two steps on the network with the implication that centrality measures for nodes of $degree=2$ would tend to cluster at approximately twice the value of nodes of $degree=1$, etc. Intermediate levels of decomposition ($100m$ and $50m$) can reduce these artefacts, thus explaining the boost to correlations; contrarily, aggressive levels of decomposition ($20m$) can have the opposite effect because the increasing proliferation of $degree=2$ nodes can exacerbate peaks in the distributions for the smallest distance thresholds. These issues tend to resolve once distance thresholds exceed $400m$.

\section{Summary}

Network analysis affords robust methods for assessing pedestrian-scale connectivity, a topic of paramount importance given that many forms of contemporary urban development continue to enforce street networks hindering pedestrianism and local access to rich assortments of land-uses. However, a confusing array of theories, methods, and considerations exist on the analysis of street networks, and it is not immediately clear which of these might be the most applicable at the pedestrian scale concerning the intensity and mix of land-uses. The preceding analysis compares correlations to mixed-uses and land-use accessibilities for a range of network centralities: weighted and unweighted, on the primal and dual network, and using shortest-path and simplest-path (angular) heuristics.

For pedestrian distances, closeness centralities offer stronger correlations to the general intensity and mix of local land-uses, both at smaller (i.e.~mixed-use high-streets) and larger (i.e.~wider mixed-use districts) pedestrian distances. Betweenness measures play a diminished role, though it could be hypothesised that betweenness is likely advantageous in contexts with sufficiently high degrees of closeness centrality, as in the case of high-streets, but is relatively meaningless in contexts with low closeness centralities, such as motorways. This observation contradicts the gist of other studies, for example, \textcite{Turner2007, Porta2009}, which have tended to favour betweenness measures, and which may be attributable to any one of numerous factors that are present to varying degrees: comparisons against traffic-counts as opposed to mixed-uses and land-use accessibilities; use of normalised closeness instead of harmonic closeness; and smaller geographic extents of analysis.

Explicitly weighted $\beta$ measures are arguably preferable to unweighted variants because they retain robust correlations against mixed-uses and land-use accessibilities, but with an explicitly configurable spatial impedance offering a substantial degree of local emphasis. The use of simplest-path measures can infer a slight advantage for correlations against mixed-uses as typified by high-streets, though these advantages are not universal and tend to dissipate for neighbourhood-scale mixed-uses and land-use accessibilities. Thus, the simplest-path premise that visual properties of the street network are core determinants in the evolution of land-uses and the broader patterns of activities in cities appears more suited to high-streets than expansive mixed-use districts where a more agglomerative and space-filling (fractal) logic is at work.

Regardless of the measure, near-maximal correlations for `high-street' mixed-uses are encountered by distance thresholds in the vicinity of $800m$ and for district-wide mixed-uses by $\approx1200m$, thus indicating little benefit to the use of larger thresholds given their increasing computational demands and the increasing effects of the Modifiable Area Unit Problem, which tends to boost correlations at larger thresholds. Conversely, centrality measures start breaking down at distance thresholds smaller than $400m$ due to increasingly exposed artefacts arising in the distributions of the measures. Within the context of pedestrian-scale mixed-uses and land-use accessibilities for Greater London, there appears to be no tangible benefit to the use of dual topologies. The greater resolution afforded by decomposition is beneficial for fine-scaled urban analysis, and moderate decomposition levels ($50m$--$100m$) may also tend to boost correlations. However, this requires consideration of the trade-off between the level of local precision required as opposed to greater computational demands.

It should be recognised that the various approaches: whether primal or dual, weighted or not, shortest or simplest, node-based or segment-based, closeness-like, betweenness-like, or otherwise more esoteric, all have some degree of relevancy, and the differences in the correlations are infrequently drastic. Therefore, either-or arguments should be avoided in recognition that multiple valid approaches exist and these may differ based on the context or the intended form of subsequent modelling. For the same reasons, models considering the impact of network centralities will benefit from dimensionality reduction to disentangle the contributions and collinearities of different methods at different scales, or else the use of non-linear machine learning methods to exploit potential nuances such as the interplay between smaller and larger thresholds, or synergies arising from the overlap of closeness, betweenness, and other forms of centrality measures. The gist of this approach is further explored in 
\ifpaper associated papers on dimensionality reduction applied to urban morphological data \parencite{Simons2021} and the detection of urban archetypes using pedestrian scale morphological information \parencite{Simons2021a}.
\else Chapters~\ref{untangling-urban-signatures} and~\ref{artificial-historical}.
\fi

\section{Acknowledgements}
\subsection{PhD}

This paper derives from the author's PhD research at the \emph{Centre for Advanced Spatial Analysis}, \emph{University College London}. The author wishes to acknowledge their PhD supervisors, Dr.~Elsa Arcaute and Prof.~Michael Batty, for their gracious support and feedback throughout the development of this work. The author takes sole responsibility for any oversights or shortcomings contained within this paper.

\subsection{Data}

\begin{flushleft}
The geographical plots and statistical figures in this document have been prepared with use of the following sources of data:\linebreak
\linebreak
\textbf{\emph{Ordnance Survey} \emph{Open Roads}}\linebreak
\emph{Contains OS data © Crown copyright and database right 2021.}\linebreak
\linebreak
\textbf{\emph{Ordnance Survey} \emph{Points of Interest} data}\linebreak
\emph{This material includes data licensed from PointX© Database Right/Copyright 2021.}\linebreak
\emph{Ordnance Survey © Crown Copyright 2021. All rights reserved. Licence number 100034829.}\linebreak
\linebreak
\textbf{\emph{UK Data Service} / \emph{Office for National Statistics} census data}\linebreak
\emph{Contains National Statistics data © Crown copyright and database right 2021.} \linebreak
\emph{Contains OS data © Crown copyright and database right (2021).}\linebreak
\end{flushleft}

\section{Citations}
\printbibliography[heading=none]{}
\end{document}